\documentclass[fleqn,usenatbib]{mnras}

\usepackage{newtxtext,newtxmath}

\usepackage[T1]{fontenc}

\DeclareRobustCommand{\VAN}[3]{#2}
\let\VANthebibliography\thebibliography
\def\thebibliography{\DeclareRobustCommand{\VAN}[3]{##3}\VANthebibliography}

\usepackage{graphicx}	
\usepackage{amsmath}	
\usepackage[separate-uncertainty=true]{siunitx}  
\usepackage{anyfontsize}

\usepackage{CJKutf8} 

\usepackage{scalerel,tikz}
\usetikzlibrary{svg.path}
\definecolor{orcidlogocol}{HTML}{A6CE39}
\tikzset{orcidlogo/.pic={
 \fill[orcidlogocol] svg{M256,128c0,70.7-57.3,128-128,128C57.3,256,0,198.7,0,128C0,57.3,57.3,0,128,0C198.7,0,256,57.3,256,128z};
 \fill[white] svg{M86.3,186.2H70.9V79.1h15.4v48.4V186.2z}
 svg{M108.9,79.1h41.6c39.6,0,57,28.3,57,53.6c0,27.5-21.5,53.6-56.8,53.6h-41.8V79.1z M124.3,172.4h24.5c34.9,0,42.9-26.5,42.9-39.7c0-21.5-13.7-39.7-43.7-39.7h-23.7V172.4z}
 svg{M88.7,56.8c0,5.5-4.5,10.1-10.1,10.1c-5.6,0-10.1-4.6-10.1-10.1c0-5.6,4.5-10.1,10.1-10.1C84.2,46.7,88.7,51.3,88.7,56.8z};
}}
\newcommand\orcidicon[1]{\href{https://orcid.org/#1}{\mbox{\scalerel*{
\begin{tikzpicture}[yscale=-1,transform shape]
\pic{orcidlogo};
\end{tikzpicture}
}{|}}}}

\newcommand{\yqy}{\begin{CJK*}{UTF8}{gkai}杨清云\end{CJK*}}
\newcommand{\cs}{c_\mathrm{s}}
\newcommand{\va}{v_\mathrm{A}}
\newcommand{\mach}{\mathcal{M}}
\newcommand{\macha}{\mathcal{M}_\mathrm{A}}
\newcommand{\rsink}{r_\mathrm{sink}}
\newcommand{\msink}{m_\mathrm{sink}}
\newcommand{\rhosink}{\rho_\mathrm{sink}}
\newcommand{\rdisc}{r_\mathrm{disc}}
\newcommand{\mdisc}{m_\mathrm{disc}}
\newcommand{\dx}{\Delta x}
\newcommand{\rhothresh}{\rho_{\mathrm{thresh}}}
\newcommand{\vk}{v_\mathrm{K}}
\newcommand{\msol}[1]{#1\,\mathrm{M}_\odot}
\newcommand{\kyr}[1]{#1\,\mathrm{kyr}}
\newcommand{\AU}[1]{#1\,\mathrm{AU}}
\newcommand{\pc}[1]{#1\,\mathrm{pc}}
\newcommand{\kms}[1]{#1\,\mathrm{km\,s^{-1}}}

\title[Protostellar disc structure and dynamics]{Protostellar disc structure and dynamics during star formation from cloud-scale initial conditions}

\author[Yang \& Federrath]{
Trey Qingyun Yang (\yqy)$^{\orcidicon{0009-0009-9033-7232}\,1}$\thanks{E-mail: \href{mailto:astro@treyyang.com}{astro@treyyang.com}; \href{mailto:trey.yang@anu.edu.au}{trey.yang@anu.edu.au}} \&
Christoph Federrath$^{\orcidicon{0000-0002-0706-2306}\,1,2}$\thanks{E-mail: \href{mailto:christoph.federrath@anu.edu.au}{christoph.federrath@anu.edu.au}}
\\
$^{1}$Research School of Astronomy and Astrophysics, Australian National University, Canberra, ACT 2611, Australia\\
$^{2}$Australian Research Council Centre of Excellence in All Sky Astrophysics (ASTRO3D), Canberra, ACT 2611, Australia
}

\date{Accepted XXX. Received YYY; in original form ZZZ}

\pubyear{\the\year{}}

\begin{document}
\label{firstpage}
\pagerange{\pageref{firstpage}--\pageref{lastpage}}
\maketitle

\begin{abstract}
The early evolution of protostellar, star-forming discs, including their density structure, turbulence, magnetic dynamics, and accretion variability, remains poorly understood. We present high-resolution magnetohydrodynamic simulations, using adaptive mesh refinement to capture detailed disc dynamics down to sub-AU scales. Starting from initial conditions derived from a molecular cloud simulation, we model the collapse of a dense core into a protostellar disc over 10,000~yr following sink particle (star) formation, achieving a maximum effective resolution of 0.63~AU. This simulation traces the evolution of the disc density, accretion rates, turbulence, and magnetic field structures. We find that the protostellar disc grows to a diameter of approximately 100~AU, with mass accretion occurring in episodic bursts influenced by the turbulence of the core from which the disc builds up. The disc is highly turbulent with a sonic Mach number of $\sim2$. Episodic accretion events within the disc cause intermittent increases in mass and magnetic energy density, resulting in an equipartition of the thermal and magnetic pressure, i.e., leading to an Alfv\'en Mach number of $\sim2$. Some regions above and below the disc mid-plane show sub-Alfv\'enic conditions with intermittent outflow activity. The disc density profiles steepen over time, following a power law consistent with observed young stellar discs and the minimum mass solar nebula. These results underscore the role of turbulence in early accretion variability and offer new insights into the physical and magnetic structure of young protostellar discs, especially with respect to their turbulent components.
\end{abstract}




\section{Introduction}

Star formation is a fundamental process shaping the structure and evolution of the interstellar medium (ISM). This process is influenced by a range of environmental factors, including gravitational forces, turbulence, magnetic fields, and feedback from new-born stars \citep{McKee_2007}. Stars form through accretion of surrounding cloud material in a protostellar disc, but also inject mass and energy back into the cloud environment via winds, jets, and radiation \citep{Stahler_2004}. Such feedback can drive turbulence in the progenitor cloud, which shapes the structure and dynamics of the ISM and may play a key role in future star formation \citep{ElmegreenScalo2004,MacLowKlessen2004,FederrathKlessen2012,HennebelleFalgarone2012,PadoanEtAl2014,BurkhartMocz2019}.

Protostellar discs extend from the central star to hundreds of AU in diameter and evolve over time-scales of millions of years \citep{Hogerheijde_2011,Najita_2018}. Over time, the gas surface density in these discs declines as the accretion rate decreases \citep{Bitsch_2015}. Recent observations of protostellar discs have made significant progress. One major discovery is that of structured distributions of gas and solids in protostellar discs, which is reshaping the field of planet formation \citep{Bae_2023}. Active research is underway to investigate the missing link between observationally derived disc properties and the formation of diverse planetary systems. Recent observations of protostellar discs around very low-mass stars have been made using ALMA, which is a challenging task due to the faintness and distance of these discs \citep{Pinilla_2021}. Observations from JWST have been able to resolve protostellar discs and probe the chemical composition of potential planet-forming regions \citep{Kospal_2023,tabone_2023}. The study of protostellar discs is important because they provide the reservoir of materials from which new planets form \citep{Zhang_2020}. Therefore, this work aims to enhance our understanding of disc structure and dynamics.

Due to the complexity of three-dimensional gas dynamics and the range of physical processes involved, computer simulations have long been used to study star formation, with increasing levels of detail over the past years. However, limits in computational resources make it impractical to simulate an entire galactic ISM, and instead simplified initial conditions have often been used, such as initially spherical cloud configurations or periodic boxes in which turbulence and magnetic fields are imposed `by hand'. These models have had great success due to their flexibility in controlling the details of the initial conditions \citep{BateBonnell1997,PriceBate2007}, enabling studies of radiation feedback \citep{Krumholz_2006,Bate2009rad,CommerconEtAl2008}, magnetic braking \citep{Mellon_2009}, the influence of the initial density profile of the cloud \citep{GirichidisEtAl2011}, effects of turbulence on the initial mass function (IMF) \citep{HaugboelleEtAl2018,NamFederrathKrumholz2021,MathewFederrathSeta2023}, disc formation in strongly magnetised cloud cores \citep{Seifried_2012,Seifried_2013}, as well as jet/outflow formation and binary star formation \citep{BossBodenheimer1979,MachidaEtAl2008,FederrathEtAl2014,Kuruwita_2017,Kuruwita_2019}. However, the outcome of these models does not arise from self-consistent initial conditions, which in reality are inherited from the large-scale molecular cloud evolution forming dense cores that are ultimately the progenitors of protostars and protostellar discs.

Therefore, in this work, we study protostellar disc formation with initial conditions inherited from a larger-scale molecular cloud simulation. We call this the `re-simulation' technique as it allows us to take any part at any time of an exiting simulation and re-run that part with enhanced resolution. The primary objective is to study disc structure and dynamics in low-mass star formation, using the most realistic initial conditions currently possible in numerical simulations. We begin by extracting a high-density core on the verge of collapse to form a protostar from a large-scale molecular cloud simulation, establishing an environment that better captures the natural turbulence, magnetic structure, and density distribution characteristic of a star-forming region.

Previous simulation works using similar inherited cloud-scale initial conditions include \citet{Kuffmeier_2017,Kuffmeier_2018,Kuffmeier_2019}, \citet{Bate_2018_Diversity}, \citet{He_2023}, and \citet{Lebreuilly_2021, Lebreuilly_2024a, Lebreuilly_2024b}. \citet{Kuffmeier_2017} modelled the formation of 6~solar-type stars during the first $\kyr{100}$ with a maximum effective resolution of 2~AU inside a $\left(\pc{40}\right)^3$ cloud. Using ideal magnetohydrodynamic (MHD) simulations, they demonstrated the accretion process to vary in time, space, and among protostars, depending on the environment. Follow-up works of infall induced accretion bursts achieved a maximum effective resolution of 0.06~AU \citep{Kuffmeier_2018} and studied the formation of companion stars in the filaments around young protostars \citep{Kuffmeier_2019}. \citet{Bate_2018_Diversity} presented statistical properties of 183~protostellar discs using smoothed particle hydrodynamics (SPH) simulations forming inside a $\left(\pc{0.404}\right)^3$ cloud, highlighting a large diversity in protostellar disc masses, radii, and orientations during the first $\kyr{90}$. \citet{He_2023} used radiation-MHD simulations to investigate the first $\sim1\,\mathrm{Myr}$ of massive, large discs around massive stars from cores of $\sim10-\msol{100}$ with a maximum effective resolution of 7.2~AU and a Jeans resolution of 10~grid cells. They similarly explored the connection between disc formation and the environment, and between turbulence and disc thickness. \citet{Lebreuilly_2021} used non-ideal MHD to examine the role of magnetic fields in the initial properties of disc populations, and later expanded on this to include the effects of radiative transfer \citep{Lebreuilly_2024a} and protostellar outflows \citep{Lebreuilly_2024b}.

Here we expand on these earlier studies by performing a re-simulation of the first $\kyr{10}$ of a single solar-type star+disc system from initial conditions provided by a $\left(\pc{2}\right)^3$ molecular cloud simulation \citep{Appel_2023}, with a particular focus on understanding the turbulent properties of the disc arising from such natural initial conditions. By adopting high-resolution adaptive mesh refinement (AMR), we capture disc details down to sub-AU scales, enabling a detailed examination of the density gradients, turbulence effects, and magnetic field configurations within the forming disc.

The study is organised as follows. In Section~\ref{sec:methods}, we define the basic numerical methods used in this work, with a particular focus on the re-simulation technique to provide initial conditions for protostellar disc formation from large-scale cloud simulations. Section~\ref{sec:results} presents our main results, discussing the density, velocity, and magnetic field structure and evolution of the disc, including its accretion, rotational and turbulent motions, magnetic field geometry, and magnetic activity. Section~\ref{sec:limitations} discusses the limitations of this work, and in Section~\ref{sec:conclusions} we present our main conclusions.

\section{Methods} \label{sec:methods}

Here we explain the technical aspects of the simulations. Sections~\ref{sec:FLASH} and \ref{sec:AMR} summarise the main simulation techniques that have also been used in previous works, while the rest of this section is devoted to introducing the re-simulation technique. Section~\ref{sec:base_sim} introduces the large-scale cloud simulation providing the initial conditions for the re-simulation, and Section~\ref{sec:resim} describes the details of the re-simulation method.

\subsection{Basic numerical methods} \label{sec:FLASH}
We use \texttt{FLASH} (version~4), a MHD simulation code with AMR capabilities \citep{BergerColella1989}, to make molecular-cloud-scale simulations computationally feasible \citep{FryxellEtAl2000,DubeyEtAl2008}. We use the HLL5R positive-definite Riemann solver \citep{WaaganFederrathKlingenberg2011} for MHD, and a multi-grid gravity solver \citep{Ricker2008}, with the system being closed by a piece-wise polytropic equation of state (EoS), which is detailed in Section~\ref{sec:EoS}.

The simulation starts with a coarse grid covering the entire computational domain. Cells are refined at each time step with higher resolution based on pre-defined thresholds and criteria (see Sec.~\ref{sec:AMR}). Conservation of physical quantities (mass, momentum, energy, etc.) is verified across coarse-fine grid interfaces as time advances \citep{Berger_1984}.

\subsubsection{Main governing equations}
We solve the three-dimensional, compressible ideal MHD equations,
\begin{equation}
    \begin{gathered}
        \frac{\partial \rho}{\partial t} + \nabla \cdot (\rho \mathbf{v}) = 0, \hfill \\
        \rho \left(\frac{\partial}{\partial t} + \mathbf{v} \cdot \nabla \right) \mathbf{v} = \frac{(\mathbf{B}\cdot\nabla)\mathbf{B}}{4\pi} - \nabla P_{\mathrm{tot}} + \rho \mathbf{g}, \hfill \\
        \frac{\partial E}{\partial t} + \left[ (E + P_{\mathrm{tot}})\mathbf{v} - \frac{(\mathbf{B\cdot v})\mathbf{B}}{4\pi} \right] = \rho \mathbf{v \cdot g}, \hfill \\
        \frac{\partial \mathbf{B}}{\partial t} = \nabla \times (\mathbf{v \times B}), \hfill \\
        \nabla \cdot \mathbf{B} = 0, \hfill
    \end{gathered}
\end{equation}
where $\rho$ is the gas density, $\mathbf{v}$ is the velocity, $P_\mathrm{tot}=P_\mathrm{thermal}+|\mathbf{B}|^2/(8\pi)$ is the total pressure from thermal and magnetic components, $\mathbf{B}$ is the magnetic field, and $E=\rho\epsilon_\mathrm{int}+\rho|\mathbf{v}|^2/2+|\mathbf{B}|^2/(8\pi)$ is the total energy density from internal, kinetic, and magnetic components. The gravitational acceleration of the gas, $\mathbf{g}$, is the sum of both the self-gravity of the gas and the contribution from sink (star) particles (see Sec.~\ref{sec:sinks}) \citep{FederrathBanerjeeClarkKlessen2010,FederrathEtAl2014},
\begin{equation}
    \begin{gathered}
        \mathbf{g} = -\nabla \Phi_\mathrm{gas} + \mathbf{g}_\mathrm{sinks}, \\
        \nabla^2 \Phi_\mathrm{gas} = 4\pi G\rho,
    \end{gathered}
\end{equation}
where $\Phi_\mathrm{gas}$ is the gravitational potential of the gas, and $G$ is the gravitational constant.

\subsubsection{Equation of state} \label{sec:EoS}
A polytropic EoS is used to obtain the pressure $P$ directly from the density $\rho$ to approximate the thermal evolution during star formation, and to close the system of MHD equations,
\begin{equation} \label{eqn:EoS}
    P_{\mathrm{thermal}} = \cs^2 \rho^\Gamma,
\end{equation}
where $\cs$ is the local sound speed, and the polytropic exponent $\Gamma$ is dependent on the local density. The polytropic EoS covers the phases of isothermal contraction, adiabatic heating during the formation of the first and second core, and the effects of $\mathrm{H_2}$ dissociation in the second collapse \citep{Larson1969, YorkeBodenheimerLaughlin1993, MasunagaInutsuka2000, OffnerEtAl2009}. The polytropic exponent is set to follow a piece-wise function,
\begin{gather}
    \Gamma = \begin{cases}
        1   & \text{for\quad} \quad\quad\; \rho \leq \rho_1 \equiv \SI{2.50e-16}{g\,cm^{-3}}, \\
        1.1 & \text{for\quad} \rho_1 < \rho \leq \rho_2 \equiv \SI{3.84e-13}{g\,cm^{-3}}, \\
        1.4 & \text{for\quad} \rho_2 < \rho \leq \rho_3 \equiv \SI{3.84e-8}{g\,cm^{-3}}, \\
        1.1 & \text{for\quad} \rho_3 < \rho \leq \rho_4 \equiv \SI{3.84e-3}{g\,cm^{-3}}, \\
        5/3 & \text{for\quad} \quad\quad\; \rho > \rho_4,
    \end{cases}
    \label{eqn:exponent}
\end{gather}
As $\Gamma$ changes from one density regime to the next, the sound speed in Eq.~(\ref{eqn:EoS}) is adjusted such that pressure and temperature are continuous functions of density. 

The initial isothermal regime ($\Gamma = 1$) approximates the thermodynamics of molecular gas of solar metallicity, over a wide range of densities \citep{WolfireEtAl1995, OmukaiEtAl2005, PavlovskiSmithMacLow2006, GloverMacLow2007a, GloverMacLow2007b, GloverFederrathMacLowKlessen2010, HillEtAl2011, HennemannEtAl2012, GloverClark2012}. In this regime, the sound speed $\cs = \SI{0.2}{km\,s^{-1}}$ and the temperature $T = \SI{11}{K}$ for molecular gas of solar composition, i.e., gas with a mean particle mass of $2.3\,m_\mathrm{H}$ \citep{KauffmannEtAl2008}, where $m_\mathrm{H}$ is the mass of a Hydrogen atom.

\subsection{Grid refinement and sink particles} \label{sec:AMR}

\subsubsection{Jeans refinement} \label{sec:jeans}
Refinement or de-refinement occurs when the change in local mass density $\rho$ causes the local Jeans length
\begin{equation} \label{eqn:jeans}
    \lambda_\mathrm{J} = \left(\frac{\pi\cs^2}{G \rho}\right)^{1/2}
\end{equation}
to drop below or exceed a pre-defined number of grid-cell lengths \citep{Jeans_1902}. All state-of-the-art numerical studies of star formation resolve the Jeans length during local collapse by more than four grid cells to avoid artificial fragmentation \citep{Truelove_1997}. However, while this minimum of 4~cells per Jeans length avoids spurious fragmentation, it is insufficient to resolve the turbulent energy content on the Jeans scale \citep{FederrathSurSchleicherBanerjeeKlessen2011}.

Therefore, we set the Jeans refinement criterion such that the local Jeans length is always resolved by 30 to 60~grid cells on all refinement levels except for the highest AMR level. This is an appropriate Jeans resolution to resolve the turbulent dynamics and minimum magnetic field amplification on the Jeans scale \citep{FederrathSurSchleicherBanerjeeKlessen2011, FederrathEtAl2014}. On the highest level of AMR, the density can accumulate and cause the local Jeans length to decrease below 30~grid cell length, but the local block cannot refine further because it has already reached the maximum allowed level of refinement. Thus, we then introduce sink (star) particles in such collapsing regions, to avoid artificial fragmentation and prohibitively small time steps, and to model star formation and accretion onto the star.

\subsubsection{Sink particle technique} \label{sec:sinks}

The complex physical processes inside a star cannot be resolved in a hydrodynamics simulation when the scale of the whole system (i.e., the cloud, core or disc scale) is much larger than that of a star. In this common situation, stars have to be modelled by a sub-resolution model such as sink particles \citep{BateBonnellPrice1995,KrumholzMcKeeKlein2004,FederrathBanerjeeClarkKlessen2010}, representing the internal properties (mass, momentum, spin, accretion rate, luminosity, etc.) of an unresolved dense gas core or star+disc system. Once created, sink particles accrete mass and momentum from their surroundings, and they can continue to interact and accrete over long periods of time. Sink particles enable the long-term evolution of systems in which localised collapse occurs, when it is impractical or unnecessary to resolve the accretion shocks at the centres of collapsing regions \citep{FederrathBanerjeeClarkKlessen2010,Gong_2013,FederrathEtAl2014}. 

Following \citet{FederrathBanerjeeClarkKlessen2010,FederrathEtAl2014}, the sink particle radius is typically set to $\rsink=2.5\,\dx$, i.e., 2.5~grid cell lengths, which is considered sufficient to capture the formation and accretion accurately, and to avoid artificial fragmentation. Requiring the Jeans lengths to be resolved by $2\,\rsink$, i.e., the diameter of a local collapsing region, defines a density threshold for sink particle creation by rearranging Eq.~(\ref{eqn:jeans}) to
\begin{equation} \label{eqn:rhosink}
\rhosink=\frac{\pi\cs^2}{4G\rsink^2}.
\end{equation}
However, when the local density exceeds this density threshold, a sink particle will not form right away. Instead, a spherical control volume with radius $\rsink$ is defined around the cell exceeding the density threshold, in which a series of checks for gravitational instability and collapse are performed. This procedure avoids spurious sink particle formation, tracing only the truly collapsing and star-forming gas.

Once a sink particle is created, it can accrete gas that exceeds $\rhosink$ inside $\rsink$, and is collapsing towards it. The excess mass is removed from the MHD system and added to the sink particle, such that mass, momentum and angular momentum are conserved by construction. Further details on the sink particle method used in this work can be found in \citet{FederrathBanerjeeClarkKlessen2010,FederrathEtAl2014}.

\begin{figure*}
    \centering
    \includegraphics[width=1\linewidth]{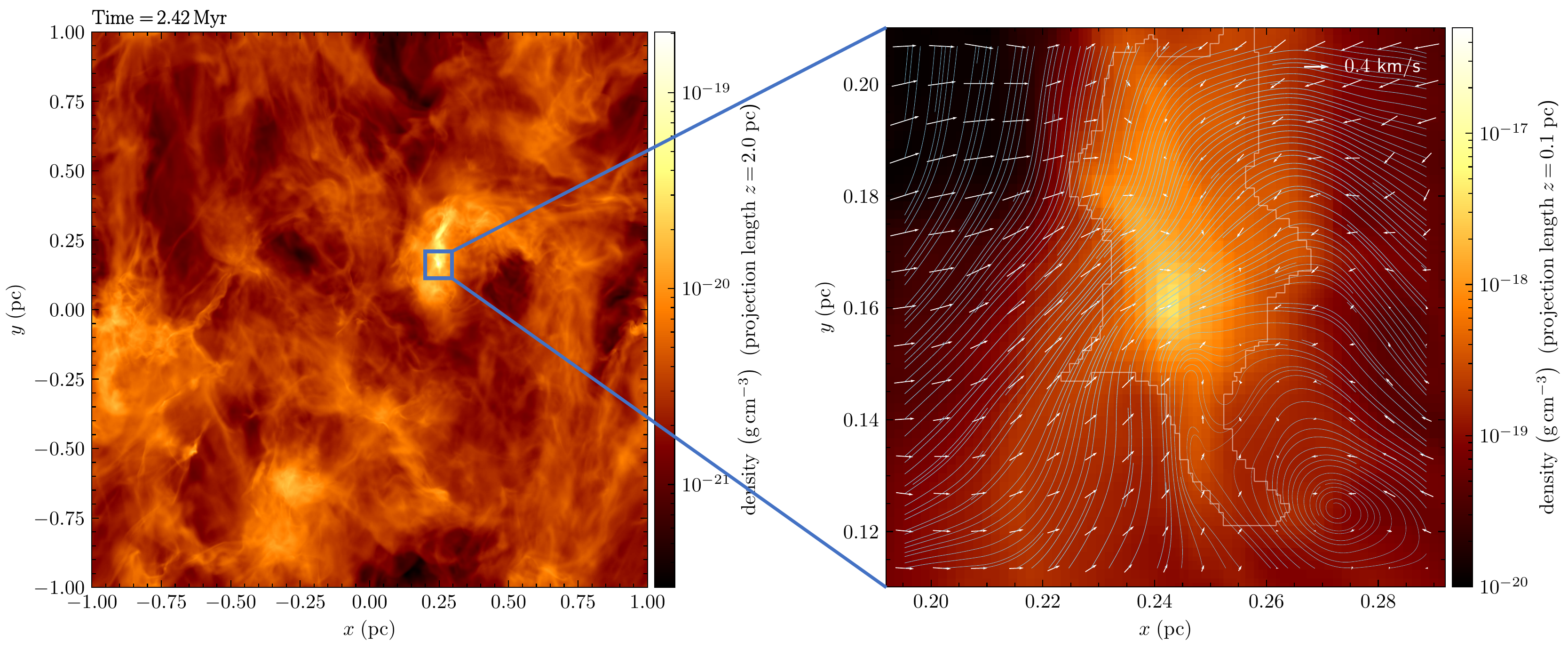}
    \caption{Column density projection of the cloud-scale base simulation (left) at the time just before the first sink particle forms at the centre of the marked square. This region serves as the initial condition for the re-simulation region (right), where velocity vectors (white arrows) and magnetic field streamlines (blue lines) are superimposed. The white contour traces the extend of the progenitor core, defined as gas above $\SI{1e-18}{g\,cm^{-3}}$ (see Sec.~\ref{sec:prep_resim_domain} and Table~\ref{tab:progenitor}).}
    \label{fig:base_sim_zoom}
\end{figure*}

\subsubsection{Resolution setting} \label{sec:resolution}
We set a maximum effective resolution of $\dx=\AU{0.63}$ for the main disc formation and evolution simulation. This is the smallest cell size at the highest level of AMR. Appendix Fig.~\ref{fig:amr} shows the AMR block structure at the end of the main run. About 90\% of the disc material is covered by the highest-resolution cells. Based on the maximum effective resolution, the sink particle radius is set to $\rsink=2.5\,\dx=\AU{1.6}$ (c.f., Sec.~\ref{sec:sinks}). We provide a resolution study in Appendix~\ref{appx:reso}, and find converging results for the main disc properties with numerical resolution, noting that full convergence requires a much higher absolute resolution, with the sink particle radius approaching the protostellar radius, which is of the order of $\mathrm{R_\odot}$, computationally intractable at this point.

\subsection{Base simulation} \label{sec:base_sim}

We use one of the simulations in the suite from \citet{Appel_2023} as the base, cloud-scale simulation from which to inherit initial conditions for the disc and star formation simulations here, i.e., to initialise the re-simulation. The molecular cloud simulation used for this is the 'GTMJR' simulation in \citet{Appel_2023}, which is an improved version of the \citet{Federrath2015} simulation at higher resolution and includes gravity, turbulence, magnetic fields, jet/outflow feedback \citep{FederrathEtAl2014}, and heating feedback \citep{MathewFederrath2020}. The left panel of Fig.~\ref{fig:base_sim_zoom} shows a 2D projection of the entire computational domain of this cloud-scale simulation at the time when the first sink particle is about to form at the centre of the region marked with the square. This cloud-scale simulation has a length of $2\,\mathrm{pc}$ on each side, periodic boundary conditions, a total cloud mass of $M=\msol{388}$, a mean density $\rho_0=\SI{3.28e-21}{g\,cm^{-3}}$, and includes mixed turbulence driving leading to a velocity dispersion of $\sigma_v = \SI{1}{km\,s^{-1}}$ and an rms Mach number of $\mathcal{M}=5$, an initial uniform magnetic field of $B = 10\,\mu G$, self-gravity, jet/outflow and heating feedback \citep[for details see][]{Appel_2023}.

While radiative heating from the star towards its environment is also included, its effects, along with the effects of jets and outflows, are not immediately relevant for the first-generation star this work focuses on, as feedback only starts after the first star has formed. However, they will be important areas of study for future works investigating interactions between sink particles -- particularly, how star formation is affected by the material contribution from previously formed stars in nearby regions. In the present work we focus exclusively on the region where the first star forms in the cloud-scale simulation as indicated by the square in the left-hand panel of Fig.~\ref{fig:base_sim_zoom}, which serves as the time and spatial region for re-simulation, shown in the right-hand panel of the figure.

\subsection{Re-simulation technique} \label{sec:resim}

\subsubsection{Determining the position and time for re-simulation}

As the large-scale base simulation evolves, we look for the first instance of the creation of a sink particle to be a candidate for re-simulation. Since we want to study the formation of a protostellar disc around a single star, it is important to ensure the absence of any other sink particles forming in the immediate region around the candidate sink particle -- an isolated star free from significant contributions from or interactions with nearby stars.

Fig.~\ref{fig:base_sim_zoom} shows the base simulation just before the formation of the first sink particle. On the left panel is the entire computational domain, with the blue box marking the $0.1\,\mathrm{pc}$ region centred around an emerging sink particle. On the right panel is a zoomed-in view of the $0.1\,\mathrm{pc}$ region. The visible pixelation indicates the length covered by the smallest grid cells in the base simulation -- the maximum effective resolution, $\dx=\AU{403}$. This is what we extract from the base simulation to provide the initial conditions for the disc and star formation re-simulation.

\subsubsection{Preparing the re-simulation domain} \label{sec:prep_resim_domain}

To prepare for re-simulation, the first step is to select a $L=0.1\,\mathrm{pc}$ box centred on the position of where the sink particle is about to form in the base simulation. This is done in the Cartesian coordinate system of the base simulation. Then, the centre-of-mass (CoM) velocity, $\mathbf{v}_\mathrm{CoM}$, of the entire re-simulation region is subtracted from every cell to effectively reset the frame of reference to that of the dense core. This is to avoid any significant bulk motion of the core with respect to the re-simulation box. Due to the different cell sizes in AMR, $\mathbf{v}_\mathrm{CoM}$ is determined by dividing the total momentum of the system by its total mass,
\begin{equation}
\mathbf{v}_\mathrm{CoM} = \frac{\sum m_i \mathbf{v}_i}{\sum m_i},
\end{equation}
where $\mathbf{v}_i$ and $m_i=\rho_i V_i$ are the velocity and mass of cell $i$, which is the product of the density ($\rho_i$) and volume ($V_i$) of the cell. The total mass contained in the re-simulation domain is $M_\mathrm{tot}=\msol{3.25}$.

\begin{table}
    \centering
    \caption{Key properties of the progenitor core defined by a density threshold $\rho_{\mathrm{prog}}=\SI{1e-18}{g\,cm^{-3}}$ in the re-simulation initial condition shown as the contoured region in the right-hand panel of Fig.~\ref{fig:base_sim_zoom}.}
    \begin{tabular}{lc}
    \hline \hline
        Parameter of progenitor core & Value \\ \hline
        Mass ($M$) & $\msol{1.34}$ \\ 
        Volume ($V$) & $\SI{1.05e51}{cm^{3}}$ \\
        Mean density ($\rho$) & $\SI{2.53e-18}{g\,cm^{-3}}$ \\ 
        Effective radius ($r_{\mathrm{eff}}=(3V/4\pi)^{1/3}$) & $\AU{4217}$ \\
        Angular momentum ($L$) & $\SI{1.65e54}{g\,cm^{2}\,s^{-1}}$ \\
        Specific angular momentum ($h$) & $\SI{6.20e20}{cm^{2}\,s^{-1}}$ \\
        Temperature ($T$) & $11\,\mathrm{K}$ \\ 
        Sound speed ($c_\mathrm{s}$) & $\kms{0.20}$ \\ 
        Velocity dispersion ($\sigma_v$) & $\kms{0.13}$ \\
        Mean magnetic field ($B$) & $44\,\mathrm{\mu G}$ \\ 
        Magnetic field dispersion ($\sigma_B$) & $46\,\mathrm{\mu G}$ \\ 
    \hline \hline
    \end{tabular}
    \label{tab:progenitor}
\end{table}

To extract further information on the initial conditions of the progenitor core, we isolate a dense region by using a density threshold $\rho_{\mathrm{prog}}=\SI{1e-18}{g\,cm^{-3}}$ and calculate a few characteristic quantities of this progenitor core, as summarised in Table~\ref{tab:progenitor}.

\subsubsection{Determining the re-simulation domain size and resolution}

The size of the re-simulation region may influence the results if too small a region is chosen. In order to obtain converged results, the re-simulation domain must be large enough to comfortably contain all the gas that ultimately forms the sink particle, its accretion disc, and any material feeding into the system. However, at the same time, we want the region to be as small as possible, to avoid unnecessary overhead, such that all available computational resources can be devoted to achieving a sufficiently high resolution of the disc, set by the maximum level of AMR. Here we chose $L=0.1\,\mathrm{pc}$ as the length for each side of the cubic re-simulation region after considering the likely diameter of the disc itself and the desired maximum effective resolution of the re-simulation, as well as the biggest region of influence from which the disc+star system is likely to accrete gas from. We conduct a box-size study (Appendix~\ref{appx:box_size}) and find that re-simulation boxes with a side length as small as $L=0.025\,\mathrm{pc}$ still produce converged results, while having increased efficiency due to being 1/4 the size of the default re-simulation size chosen here. We chose a somewhat larger box size than necessary here ($L=\pc{0.1}$), in case this simulation were to be evolved to much later times in a follow-up study, in which case material from further away could in principle reach the disc+star system.

The re-simulation uses inflow/outflow boundary conditions. However, the choice of boundary condition is inconsequential, as the re-simulation domain size was selected to be sufficiently large to avoid any boundary effects to propagate to the forming star+disc system. The resolution of the re-simulation is set to $\dx=\AU{0.63}$ and $\rsink=2.5\,\dx=\AU{1.6}$ (c.f., Sec.~\ref{sec:resolution}). We also provide a resolution study in Appendix~\ref{appx:reso}.

\section{Results} \label{sec:results}

\begin{figure*}
    \centering
    \centering
    \includegraphics[width=0.95\linewidth]{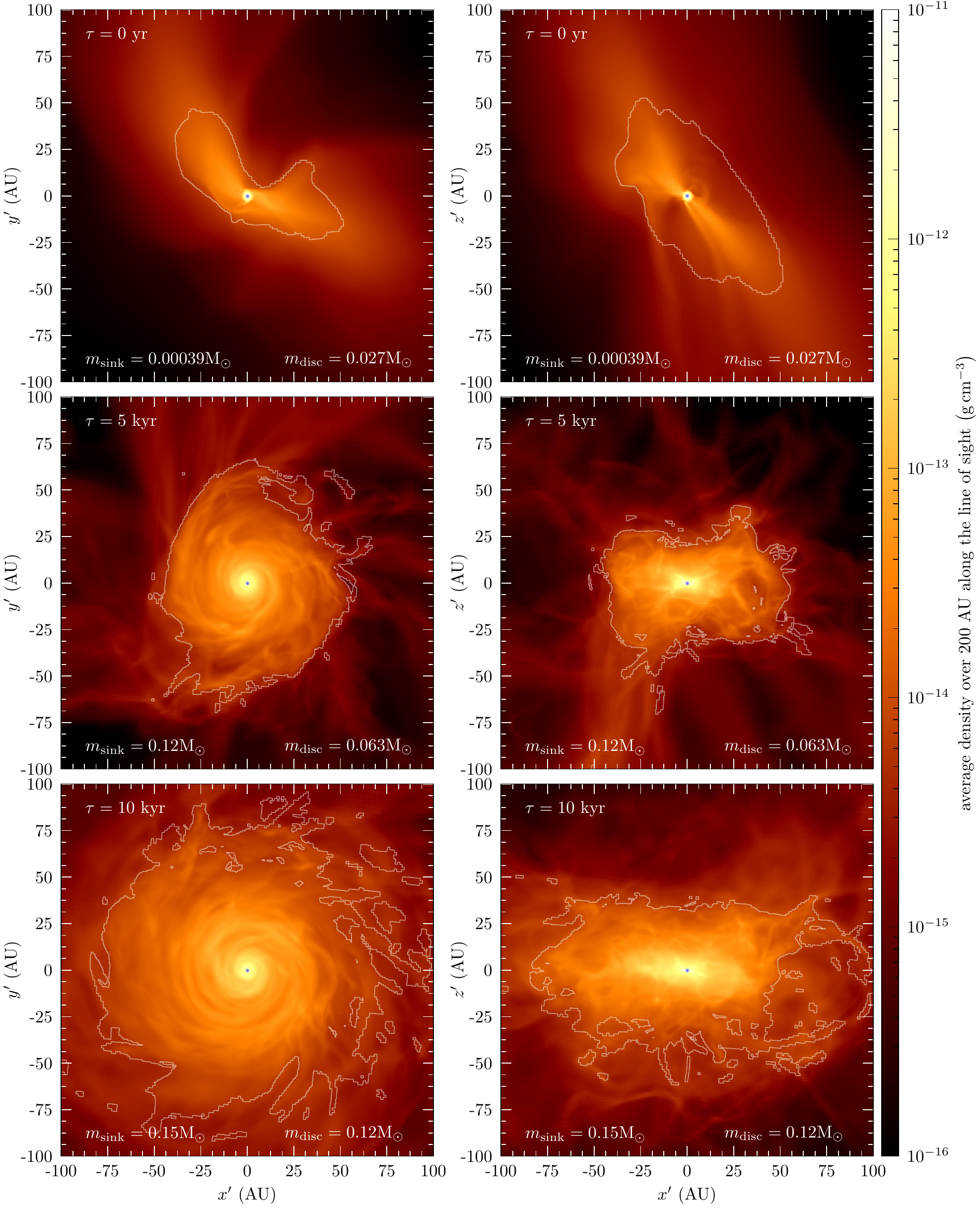}
    \caption{Column-density projections of the disc region at $\tau=0$, $5$, and $\kyr{10}$ after sink formation (from top to bottom), zoomed-in to show 200~AU on each axis. The re-simulation coordinates $(x,y,z)$ have been transformed into translated and rotated coordinates $(x',y',z')$, such that the disc's angular momentum vector is along $z'$, and $(x',y',z')=(0,0,0)$ is the location of the sink particle (shown as a blue asterisk). The contour lines trace the extent of the disc material, as defined by $\rhothresh=\SI{3.847e-14}{g\,cm^{-3}}$. The left and right columns show the disc face-on and edge-on, respectively. We see how an initially filamentary shaped accretion stream evolves into a disc, feeding via several accretion streams at later times. The disc and its surroundings display complex turbulent structures as well as spiral arms. The edge-on view reveals that the disc has a fairly large scale height considering its radial extent, as it is undergoing highly dynamic accretion from a turbulent environment during its young evolutionary stages.}
    \label{fig:evo_rotation}
\end{figure*}

\begin{table}
    \setlength{\tabcolsep}{1.5pt}
    \centering
    \caption{Similar to Table~\ref{tab:progenitor}, but for the protostellar disc, defined by a density threshold $\rhothresh=\SI{3.847e-14}{g\,cm^{-3}}$, at a protostellar age of $\tau=\kyr{10}$.}
    \begin{tabular}{lc}
    \hline \hline
        Parameter of protostellar disc & Value \\ \hline
        Mass ($M$) & $\msol{0.12}$ \\ 
        Volume ($V$) & $\SI{1.20e45}{cm^{3}}$ \\
        Mean density ($\rho$) & $\SI{2.09e-13}{g\,cm^{-3}}$ \\ 
        Effective radius ($r_{\mathrm{eff}}$) & $\AU{44}$ \\
        Angular momentum ($L$) & $\SI{2.06e52}{g\,cm^{2}\,s^{-1}}$ \\
        Specific angular momentum ($h$) & $\SI{8.23e19}{cm^{2}\,s^{-1}}$ \\
        Mean temperature ($T$) & $21\,\mathrm{K}$ \\ 
        Mean sound speed ($c_\mathrm{s}$) & $\kms{0.28}$ \\ 
        Velocity dispersion ($\sigma_{v_r},\,\sigma_{v_\varphi},\,\sigma_{v_{z'}}$) & $(0.37,\,0.30,\,0.33)\kms{}$ \\
        Mean magnetic field ($B_r,\,B_\varphi,\,B_{z'}$) & $(-4.3,\,150,\,-1.1)\,\mathrm{mG}$ \\ 
        Magnetic field dispersion ($\sigma_{B_r},\,\sigma_{B_\varphi},\,\sigma_{B_{z'}}$) & $(39,\,55,\,43)\,\mathrm{mG}$ \\ 
    \hline \hline
    \end{tabular}
    \label{tab:disc}
\end{table}

Here we present the main results of the re-simulation of protostellar disc formation. We run the simulation to 10,000~yr after sink (protostar) formation ($\tau=\kyr{10}$) to allow the accretion disc sufficient time to develop and evolve.

We begin by defining the disc material in Section~\ref{sec:discdef}, followed by Section~\ref{sec:density} focusing on the density statistics of the disc, analysing the time evolution of the disc mass and size, radial profiles, and disc scale height. Section~\ref{sec:kinematics} analyses the disc kinematics, including the velocity distribution and turbulent velocity fluctuations. Section~\ref{sec:magnetics} discusses the magnetic field structure, including the mean and turbulent field components. The global disc properties at $\tau=\kyr{10}$ are listed in Table~\ref{tab:disc}.

\subsection{Definition of disc material} \label{sec:discdef}

The disc material is defined by a density threshold of $10^{10}\,\mathrm{cm^{-3}}$, which corresponds to $\rhothresh=\SI{3.847e-14}{g\,cm^{-3}}$. The resulting disc material is primarily confined to a $\sim\AU{200}$ region around the sink particle, shown by the contours in Fig.~\ref{fig:evo_rotation}. Appendix Fig.~\ref{fig:amr} shows the AMR block structure superimposed on the bottom panels of Fig.~\ref{fig:evo_rotation}.

\subsection{Disc density statistics} \label{sec:density}

\subsubsection{Structure and morphology}

Fig.~\ref{fig:evo_rotation} shows face-on and edge-on views of the disc at three points in time: when the sink particle forms ($\tau=0\,\mathrm{yr}$), 5,000~yr after sink formation ($\tau=\kyr{5}$), and 10,000~yr after sink formation ($\tau=\kyr{10}$). Face-on views are obtained by rotating the disc such that the total angular momentum vector of the disc aligns with the line of sight, while edge-on views are rotated such that the angular momentum vector is perpendicular to the line of sight.

We see that the disc grows in both diameter and thickness. It quickly settles into a state of rotation, accretion, and spiralling. An animation of the disc evolution is available online as supplementary material. In the face-on views of the disc at $\tau=\kyr{5}$ and $\tau=\kyr{10}$, we can see spiral features, which can be attributed to instabilities in young discs \citep{Cossins_2009, Forgan_2011, Hall_2018, Bethune_2021}. Overall, this system is highly dynamic, with several accretion streams feeding material from the turbulent progenitor core onto the disc.

\subsubsection{Time evolution of disc mass and size} \label{sec:12evo}

\begin{figure}
    \centering
    \includegraphics[width=\linewidth]{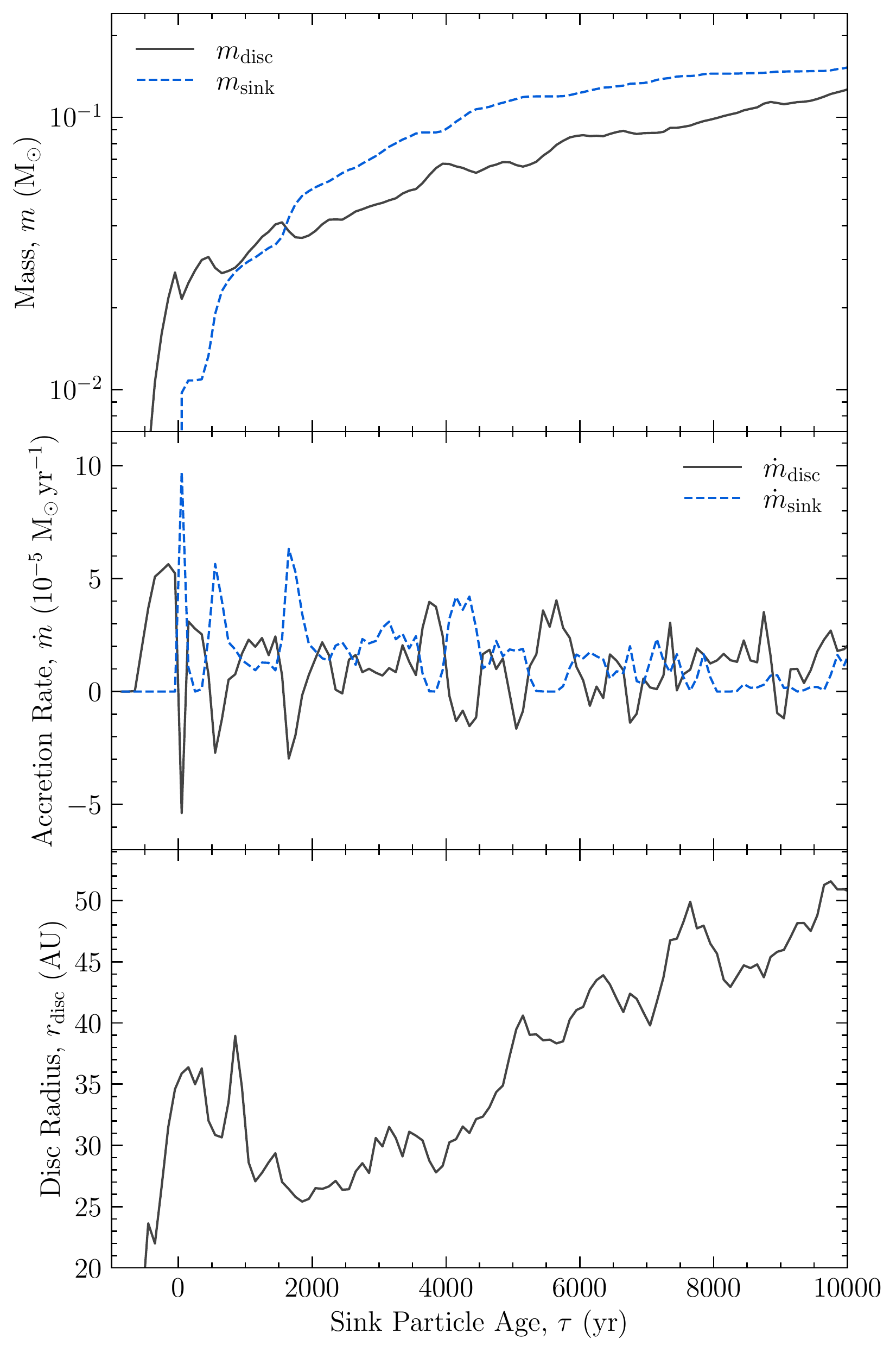}
    \caption{Time evolution of sink particle mass and disc mass (top), accretion rates (middle), and disc radius (bottom). The sink particle mass maintains an upwards trend through accretion, however, instead of a continuous mass intake from the disc, there are episodic fast accretion activities followed by slowdown and even brief pauses. The disc maintains general growth as well despite periods of mass/size decreasing, indicating mass being fed from the disc into the sink particle. The disc radius increases overall, but also shows fluctuations on time scales of a few hundred years, suggesting a correlation with the episodic accretion rate.}
    \label{fig:12_time_evo}
\end{figure}

Fig.~\ref{fig:12_time_evo} shows the time evolution of the sink particle mass and the disc mass, their respective rates of change, and the disc radius. The disc radius, $\rdisc$, is defined by the distance (cylindrical radius) from the sink particle that contains 80\% of the cumulative mass of all disc material in the cylinder (disc mass is defined by the density threshold introduced in Sec.~\ref{sec:discdef}). Following \citet{Bate_2018_Diversity}, we reject the remaining 20\% of disc material in the definition of $\rdisc$, because in the outer disc, mass diffuses and flares out to large radii, which does not reflect the concentration of material in the disc itself. It should be noted that while some studies take velocity profiles into consideration in the definition of disc size, we follow the definition in \citet{Bate_2018_Diversity}.

Overall, we see a steady accretion of mass for both disc and sink particle, and $\msink$ seems to asymptote as it approaches the end of the 10,000~yr simulation period. Although it is now already firmly in the realm of being a low-mass star with mass $\msink=\msol{0.15}$, we expect it to continue growing for a longer time, until it reaches a quiescent state typical for low-mass stars a few hundred~kyr after sink formation \citep{Evans_2009}. The disc-to-star mass ratio remains close to unity, and is consistent with the findings of \citet{Bate_2018_Diversity}.

The mass accretion rate of the sink and the disc appear anti-correlated, as high values of $\dot{m}_{\mathrm{sink}}$ are accompanied by low values of $\dot{m}_{\mathrm{disc}}$. One can roughly draw a line at $\dot{m}\sim10^{-5}\,\mathrm{M_\odot\,yr^{-1}}$, which intercepts the inflection points of $\msink$ and $\mdisc$ as a `baseline' accretion rate of the system. This shows the steady accretion of mass for both disc and sink. However, the sink particle accretes mass from the inner parts of the disc in episodes of high accretion rates, which leads to a corresponding drop in the disc mass. The disc mass gets replenished and increases by accreting material from the surrounding dense cloud core.

It has long been demonstrated observationally that early disc accretion is highly episodic, and the accretion rate can range from $10^{-7}$ to a few times $10^{-4}\,\mathrm{M_\odot\,yr^{-1}}$ for low-mass stars \citep{Hartmann_1996_FU_Ori}. This is consistent with our disc, and the episodic nature of accretion in particular is a caused by the turbulent initial conditions inherited from the cloud-scale simulation, as opposed to idealised initial conditions (e.g., with spherical symmetry and/or a uniform gas distribution). Similar to the results from \citet{Kuffmeier_2017}, we find an accretion rate profile that is slowly decreasing with time.

Finally, the bottom panel of Fig.~\ref{fig:12_time_evo} shows that the disc grows to a radius of $\sim\AU{50}$ with fluctuations on $\sim100\,\mathrm{yr}$ time-scales, similar to the episodic accretion time-scales.

\subsubsection{Disc density profiles} \label{sec:densityProfiles}

\begin{figure}
    \centering
    \includegraphics[width=\linewidth]{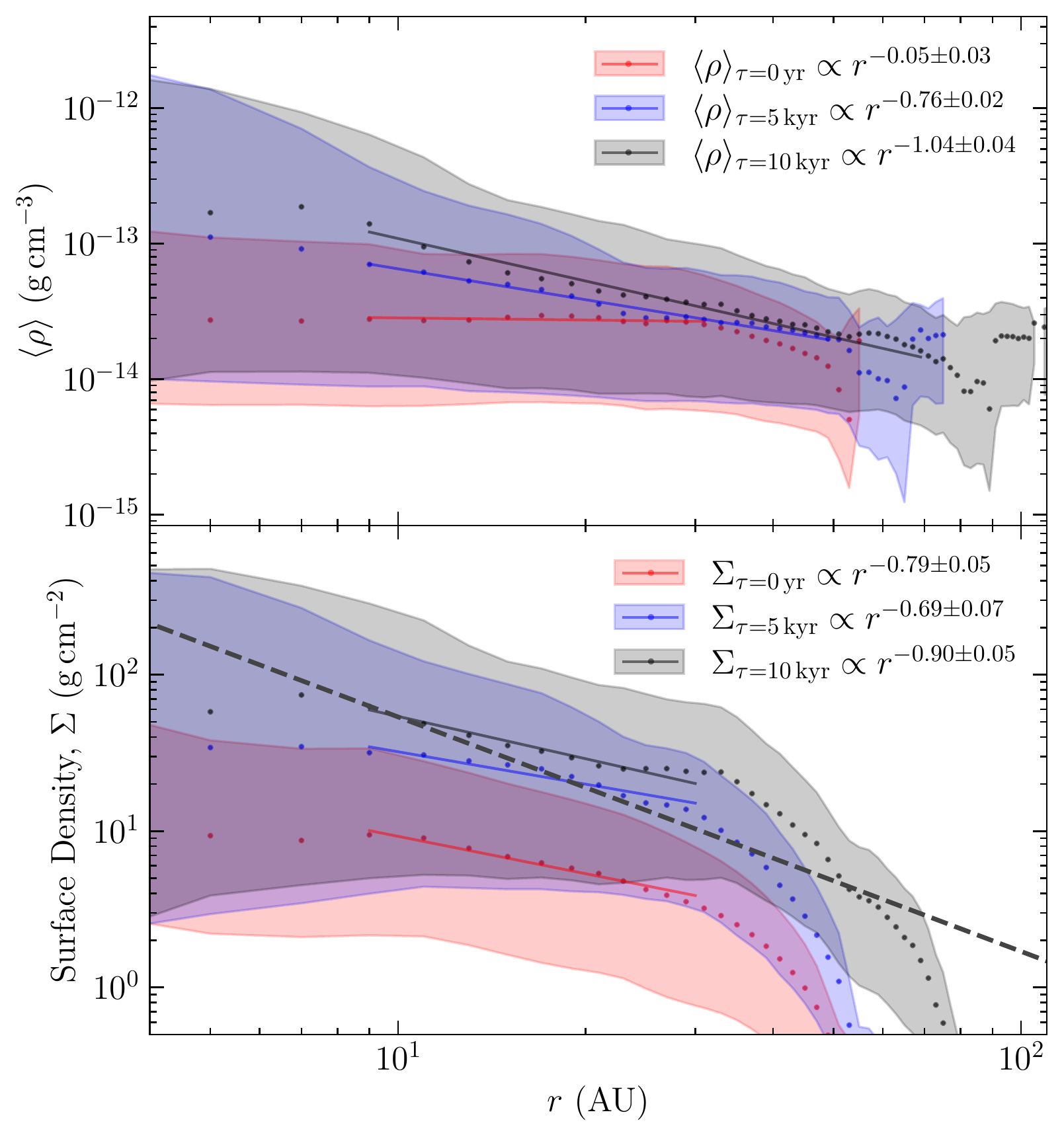}
    \caption{Radial profiles of the disc density $\langle\rho\rangle$ (top) and surface density $\Sigma$ (bottom) at $\tau=0$ (red), $5$ (blue), and $\kyr{10}$ (black), shown by the solid dots with 16th to 84th percentile ranges shown as the shaded regions. The solid straight lines show power-law fits with Eq.~(\ref{eqn:power}), where the fitted power-law exponents are denoted in the legend and listed in Table~\ref{tab:densityPowerFitting}. In the bottom panel, the solid straight line marks the MMSN relation of $\Sigma=1700\,\mathrm{g\,cm^{-2}} \left( r/\mathrm{AU} \right)^{-1.5}$.}
    \label{fig:densProfile}
\end{figure}

\begin{table}
    \setlength{\tabcolsep}{3.8pt}
    \centering
    \caption{Power-law fit parameters based on Eq.~(\ref{eqn:power}) of the radial density profiles shown in the top (for volumetric density $\rho$) and bottom (for column density $\Sigma$) panels of Fig.~\ref{fig:densProfile}.}
    \label{tab:densityPowerFitting}
    \begin{tabular}{cllll}
    \hline\hline
    \multicolumn{1}{c}{$\tau\,\mathrm{(kyr)}$} & \multicolumn{1}{c}{$\rho_0\,\mathrm{(g\,cm^{-3})}$} & \multicolumn{1}{c}{$\alpha$ (for $\rho$)} & \multicolumn{1}{c}{$\Sigma_0\,\mathrm{(g\,cm^{-2})}$} & \multicolumn{1}{c}{$\alpha$ (for $\Sigma$)} \\ \hline
    0 & $\SI{3.2\pm0.3e-14}{}$ & $0.05\pm0.03$ & $\phantom{0}58\pm7$ & $0.79\pm0.05$ \\
    5 & $\SI{3.7\pm0.2e-13}{}$ & $0.76\pm0.02$ & $157\pm30$ & $0.69\pm0.07$ \\
    10 & $\SI{1.2\pm0.1e-12}{}$ & $1.04\pm0.04$ & $430\pm60$ & $0.90\pm0.05$ \\
    \hline\hline
    \end{tabular}
\end{table}

We describe the disc geometry within a cylindrical coordinate system, with the centre of the disc at the origin, and $r$, $\varphi$, and $z'$ being the radial, azimuthal, and vertical directions, respectively. The top panel in Fig.~\ref{fig:densProfile} shows the radial density profile of the disc at three points in time: $\tau=0$, $5$, and $\kyr{10}$. The solid dots show the median, while the shaded region encloses the interval from the 16th to the 84th percentile. The latter shows the significant level of turbulent density fluctuations as the disc accretes from its surrounding. On average, however, we see that the density increases towards the centre at later times, while the density at the disc radius roughly stays the same as the disc grows in radius. The overall densities for $r\sim\AU{10}$ agree with those in \citet{Kuffmeier_2017}.

Overall, we see that the relationship between density and radius can be described with a power law,
\begin{equation} \label{eqn:power}
    \langle \rho \rangle = \rho_0 (r/\mathrm{AU})^{-\alpha},
\end{equation}
where $\langle \rho \rangle$ is the density averaged over all $\varphi$ and $z'$ at a given radius, $\rho_0$ is density at $r=\AU{1}$, and $\alpha$ is the power-law exponent.

Fitting is performed for $r \geq \AU{9}$ to exclude cells too close to the sink particle and cells affected by numerical diffusion. The $r = \AU{9}$ scale corresponds to $\sim 28\,\dx$ in diameter, very close to where we expect numerical effects to be minimal or absent, i.e., for scales $\gtrsim30\,\dx$ \citep{KitsionasEtAl2009,FederrathDuvalKlessenSchmidtMacLow2010,SurEtAl2010,FederrathSurSchleicherBanerjeeKlessen2011,MalvadiFederrath2025}. The fit parameters are listed in Table~\ref{tab:densityPowerFitting}. As expected, $\rho_0$ and $\alpha$ both increase as mass is accreted over time and the density profiles steepen. At $\tau=0$, $\alpha=0.05\pm0.03$, i.e., the density profile is practically flat, while at $\tau=5$ and $\kyr{10}$, $\alpha=0.76\pm0.02$ and $1.04\pm0.04$, respectively, quantifying the steeping over time, until an exponent of $\alpha\sim1$ is reached.

The bottom panel of Fig.~\ref{fig:densProfile} shows the surface density, $\Sigma=\int\rho\,dz'$, as a function of disc radius. We see that the relationship between surface density and radius can also be described with a power law, equivalent to Eq.~(\ref{eqn:power}) but for $\Sigma$ instead of $\rho$. As for $\rho$ in the top panel, we fit $\Sigma$ with this power-law function and list the fit parameters in Table~\ref{tab:densityPowerFitting}. Overall, we find surface density profiles to be shallower than the minimum mass solar nebula (MMSN) (black dashed line in Fig.~\ref{fig:densProfile}, $\Sigma=1700\,\mathrm{g\,cm^{-2}} \left( r/\mathrm{AU} \right)^{-1.5}$, which describes the lowest mass required to form the present-day Sun and the planets in our solar system, assuming negligible effects of magnetic field \citep{Weidenschilling_1977,Hayashi_1981}. At $\tau=\kyr{10}$, the disc has not yet reached a mature state, as suggested by the continuous increase in disc mass (see Fig.~\ref{fig:12_time_evo}). Speculating from the continuous increase in $\alpha$ with time, and the 50-kyr discs presented by \citet{Kuffmeier_2017}, the disc can evolve to be a closer match to the MMSN model in the future as more mass is accreted into the inner disc region. However, it is worth noting that the strong magnetic effects reduce the surface density in the inner regions of the disc, causing angular momentum transport and magnetic braking \citep{Wurster_2018}.

\subsubsection{Disc scale height} \label{sec:scale_height}

\begin{figure*}
    \centering
    \includegraphics[width=.95\linewidth]{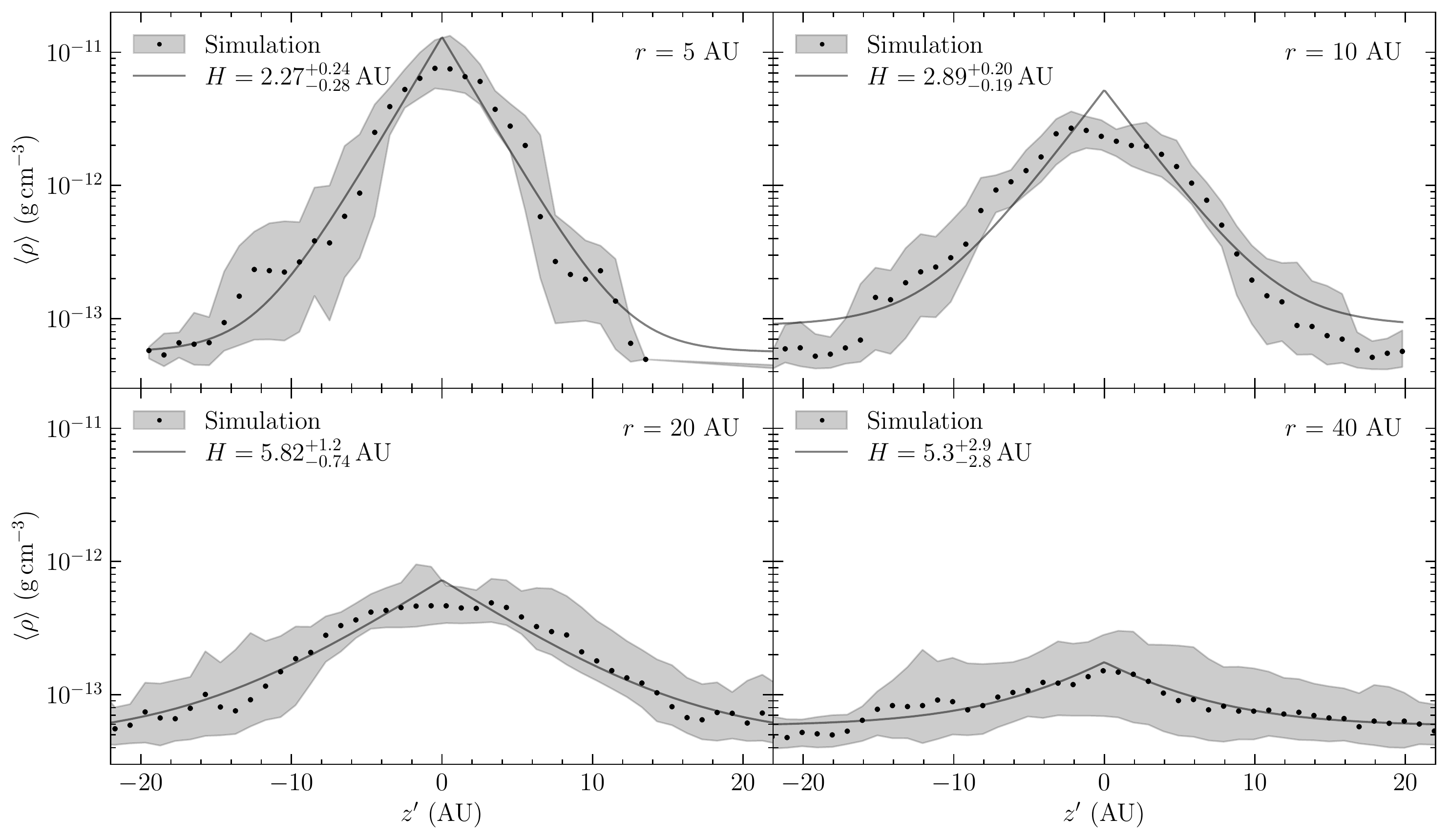}
    \caption{Vertical density profiles of the disc at different radii, $r=5$, $10$, $20$, and $\AU{40}$, at $\tau=\kyr{10}$. The median values of the density profiles are shown as the black dots with the 16th to 84th percentile ranges shown as the shaded regions. The solid lines show exponential decay fits with Eq.~(\ref{eqn:Hfit}), and the resulting disc scale height, $H$, is denoted in the legends.}
    \label{fig:Hseveral}
\end{figure*}

\begin{figure}
    \centering
    \includegraphics[width=.95\linewidth]{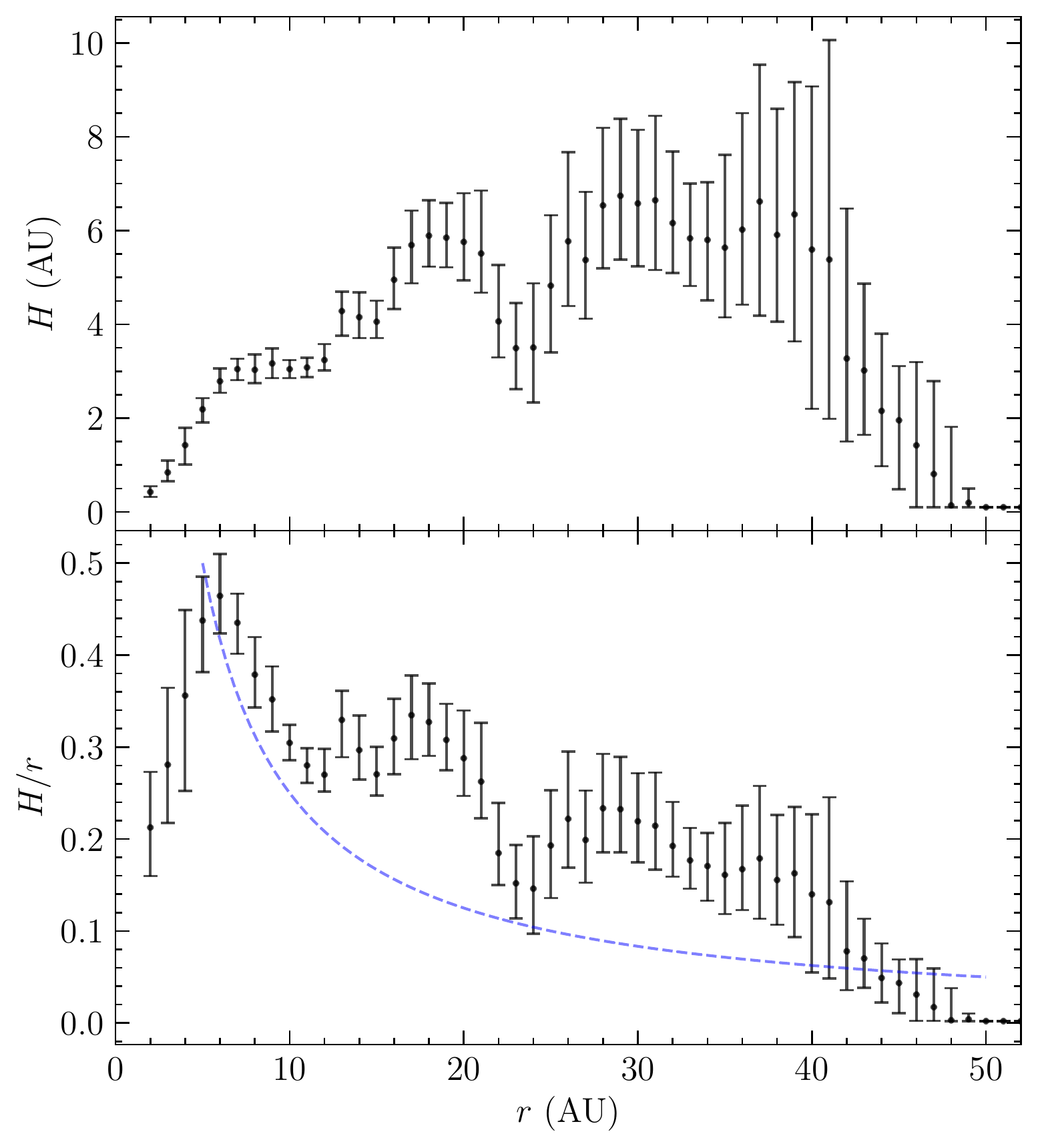}
    \caption{The disc scale height $H$ (top panel), fitted via Eq.~(\ref{eqn:Hfit}) as a function of disc radius. We find a general trend of increasing $H$ as $r$ increases, up to $r \sim \AU{40}$, indicating some flaring of the disc. The aspect ratio $H/r$ (bottom panel) is decreasing more slowly than $\propto r^{-1}$ (the aspect ratio if $H$ were constant; shown as the blue dashed line for comparison), quantifying the amount of flaring. Moreover, the disc is geometrically thick with $H/r\sim0.5-0.1$ for $r\sim5-40\,\mathrm{AU}$, due to the high levels of turbulence.
    }
    \label{fig:Hprofile}
\end{figure}

The disc scale height is a characteristic quantity that describes the vertical structure of the accretion disc. To calculate the scale height at a specific radius, the density is binned along $z'$ and fitted with an exponential decay function,
\begin{equation} \label{eqn:Hfit}
    \langle \rho \rangle = \rho_0 e^{-|z'|/H}+\rho_\mathrm{BG},
\end{equation}
where $\langle \rho \rangle$ is the density averaged over all $\varphi$ at a given radius $r$, $\rho_0$ is the mid-plane density (at $z'=0$), $H$ is the scale height, and $\rho_\mathrm{BG}$ is the background density as $|z'|\to\infty$.

Unless stated otherwise, from this point forward, we focus our discussion on the disc at $\tau=\kyr{10}$, i.e., the most evolved time of the simulation that we were able to reach with the current computations. Fig.~\ref{fig:Hseveral} shows the volume-weighted density distribution and exponential decay fits with Eq.~(\ref{eqn:Hfit}) at $r=5$, $10$, $20$, and $\AU{40}$. The black dots show the median density, while the shaded regions enclose the interval from the 16th to 84th percentile. The density profiles plateau below the fits around $z'=0$, because the sink particle accretes mass in excess of $\rhosink$ from cells within its radius $\rsink=\AU{1.6}$ (see Eq.~\ref{eqn:rhosink}). The density profiles transition into the background density $\rho_\mathrm{BG}\sim\rhothresh$ as $|z'|\to\infty$. We see that the fits provide a reasonably good approximation to the simulation profiles.

Among the scale heights at the presented radii, we find an increase in $H$ as $r$ increases. At smaller radii, the density changes more quickly with height above the disc mid-plane. We fit for scale heights along the entire disc to produce a radial profile of $H$. The result is shown in the top panel of Fig.~\ref{fig:Hprofile}. We see the scale height continues to increase with radius, reaching a maximum of $H\sim\AU{6}$ around $r\sim\AU{40}$, indicating a flaring disc, followed by a rapid decrease with radius beyond that. While our disc+star system reaches an age $\tau=\kyr{10}$, this age is still considered very early in the evolution of a low-mass star. It is therefore no surprise to find our disc flaring at large radii, as turbulent material from the dense ambient core falls onto the disc, continuously perturbing the disc's vertical structure.

In the bottom panel of Fig.~\ref{fig:Hprofile}, the ratio of $H/r$ shows a steady and constant decrease in the entire disc after $r\geq\AU{5}$. Closer to the sink particle, i.e., $r<\AU{5}$, we observe a distinct region where $H/r$ increases with $r$, likely associated with the active accretion region onto the sink particle (see Sec.~\ref{sec:vMorphology}). Following \citet{McKee_2007}, we can identify the boundary between inner and outer disc at $r\sim\AU{40}$, defined by where $H/r$ drops below 0.1, however, a prominent feature still remains at $\AU{20}\leq r \leq\AU{30}$, where a disruption in the continuity of the disc density profile likely indicates an intermittent, bursty, turbulent accretion event at that radius range and particular time in the disc evolution. Overall, $r/H$ stays relatively high at all radii in the disc, showing that the disc is geometrically thick, which is a consequence of its dynamic turbulent state during its young accretion phase. A similar connection between turbulence and disc thickness was also explored by \citet{He_2023}.

The disc flaring and geometrical thickness is visually apparent from a cross-section view of the disc in the right-hand panel of Fig.~\ref{fig:slice}, showing the highly perturbed, turbulent density structure of the disc. The sharp decrease in $H$ between $r=\AU{20}$ and $r=\AU{30}$ is our first indication of a gap between the inner and outer disc at this particular snapshot in time. The existence of an optically thin gap between optically thick disc radial segments has been detected observationally, and \citet{Espaillat_2007, Andrews_2009} have linked the gap in mature (a few Myr-old) discs to planet formation. Here, however, it is associated with the episodic nature of turbulent accretion and disc perturbation during the young evolutionary stages of the disc.

\subsection{Disc kinematics and velocity statistics} \label{sec:kinematics}

The preceding discussion of the disc density statistics has pointed towards a disc+star system that is characterised by turbulent fluctuations and episodic accretion events. We therefore proceed now to analyse the kinematic properties of the disc, with a particular focus on its turbulent velocity fluctuations.

\subsubsection{Structure and morphology}
\label{sec:vMorphology}

\begin{figure*}
    \centering
    \centering
    \includegraphics[width=.95\linewidth]{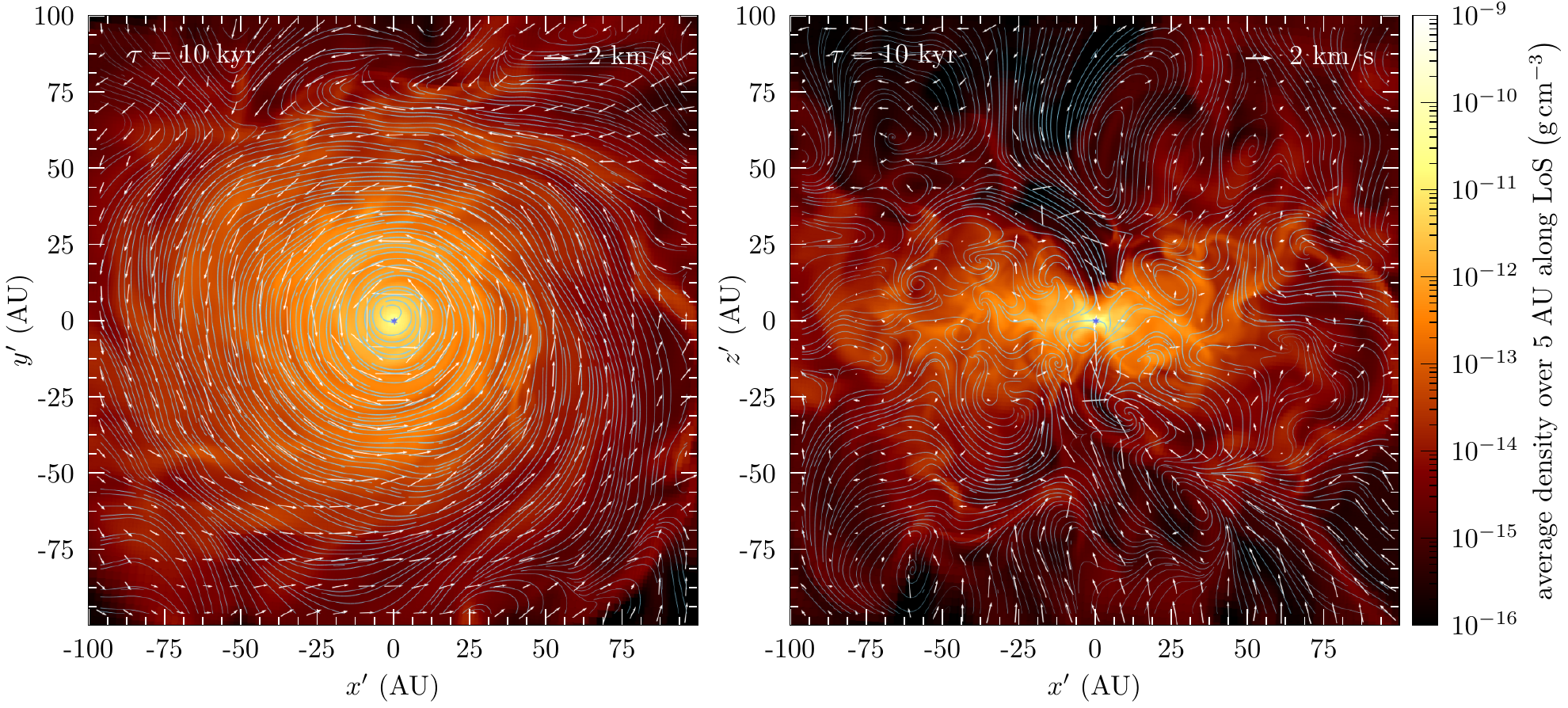}
    \caption{Density slices of the disc at $\tau=\kyr{10}$, face-on (left) and edge-on (right). The blue asterisk marks the location of the sink particle. This view captures a $\AU{5}$-thick `cross section' of the disc around the sink particle. White arrows represent velocity vectors, which follow the direction of gas motion, and their lengths is proportional to the magnitude of the velocity. Magnetic field streamlines are shown in blue.}
    \label{fig:slice}
\end{figure*}

Fig.~\ref{fig:slice} shows a cross section of the disc at $\tau=\kyr{10}$ with velocity vectors representing the gas motion. These projections represent the average density and velocity over 5~AU along the line of sight to avoid obfuscation from outer disc material. The face-on view (left panel) shows prevalent rotation in the counter-clockwise direction. Radial motions are present and correspond to accretion flows through the disc, but the azimuthal velocity component dominates strongly over the radial velocity component. For the vast majority of the disc mid-plane, rotational motion is of the order of 2~km/s. The rotation of the disc is sped up from converted gravitational potential energy. The edge-on view (right-hand panel) shows clear flaring in the outer disc out to $r\sim\AU{40}$ (as quantified in previous Fig.~\ref{fig:Hprofile}), where the thickness of the disc is of the order of its diameter. In the regions above and below the sink particle along the $z'$-axis ($\AU{-10} \lesssim x' \lesssim\AU{10}$), the velocity vectors show material being accreted preferentially into the inner disc instead of the sink particle directly, except for regions very close to the sink particle.

\subsubsection{Keplerian motion analysis} \label{sec:keplerian}

\begin{figure}
    \centering
    \includegraphics[width=\linewidth]{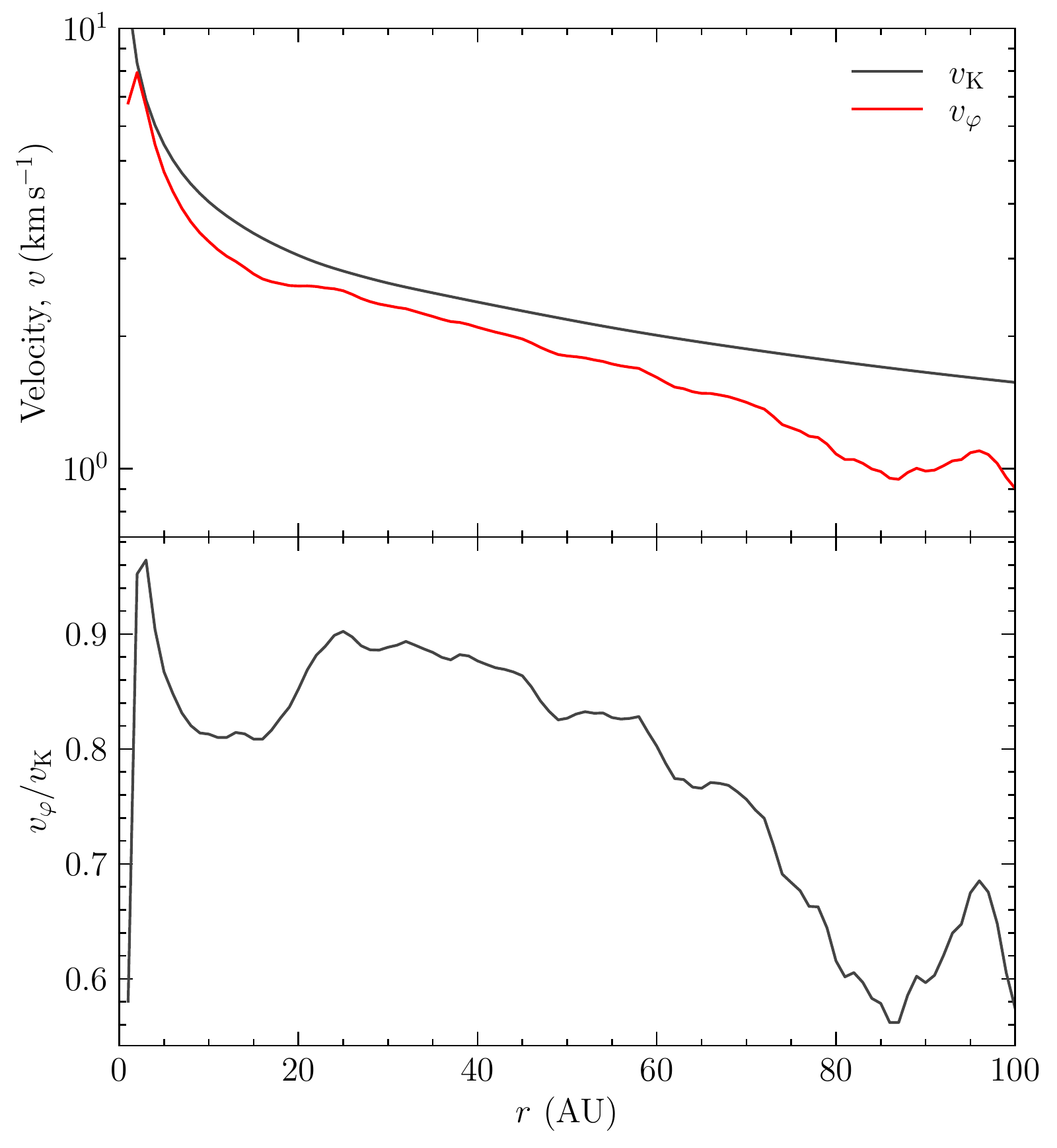}
    \caption{Keplerian analysis of the disc at $\tau=\kyr{10}$, showing the Keplerian velocity $v_\mathrm{K}$ with the rotational velocity $v_\varphi$ (top), and $v_\varphi/\vk$ (bottom) as a function of disc radius. The rotation curve, $v_\varphi$, is close to Keplerian near the centre ($r\sim\AU{5}$), while being mildly sub-Keplerian throughout the rest of the disc ($v_\varphi/\vk\sim0.8-0.9$), and $v_\varphi/\vk\sim0.6-0.7$ beyond the disc radius ($r\sim\AU{50}$).}
    \label{fig:keplerian}
\end{figure}

To further understand the overall kinematics of the disc, we perform a Keplerian motion analysis at $\tau = \kyr{10}$ by first finding the cumulative disc mass $M(r) = \int dz' \int d\varphi \int_{0}^{r} d\tilde{r}\;m(\tilde{r},\varphi,z')$ as a function of disc radius, and add the sink particle mass ($\msink$); then calculating the Keplerian velocity as,
\begin{equation}
    \vk(r) = \left( \frac{G[M(r)+\msink]}{r} \right)^{1/2}.
\end{equation}
Finally, we compute the ratio between the rotational velocity and the Kepler speed as $v_\varphi/\vk$, in order to quantify whether or not the disc follows Keplerian rotation. Fig.~\ref{fig:keplerian} shows the results. We see that $v_\varphi$ is close to Keplerian just before the gas is accreted onto the sink particle (at $r\sim\AU{5}$), while $v_\varphi/\vk$ shows mildly sub-Keplerian motion with $v_\varphi/\vk\sim0.8-0.9$ throughout the rest of the disc out to the disc radius ($r\sim\AU{50}$). This is followed by a drop in $v_\varphi/\vk$ to $\sim0.6-0.7$ immediately outside the disc, reflecting accretion. These results are broadly consistent with other studies \citep{Nordlund_2014,Marchand_2020,Mayer_2025}.

\subsubsection{2D maps of velocity and velocity dispersion}

\begin{figure*}
    \centering
    \includegraphics[width=\linewidth]{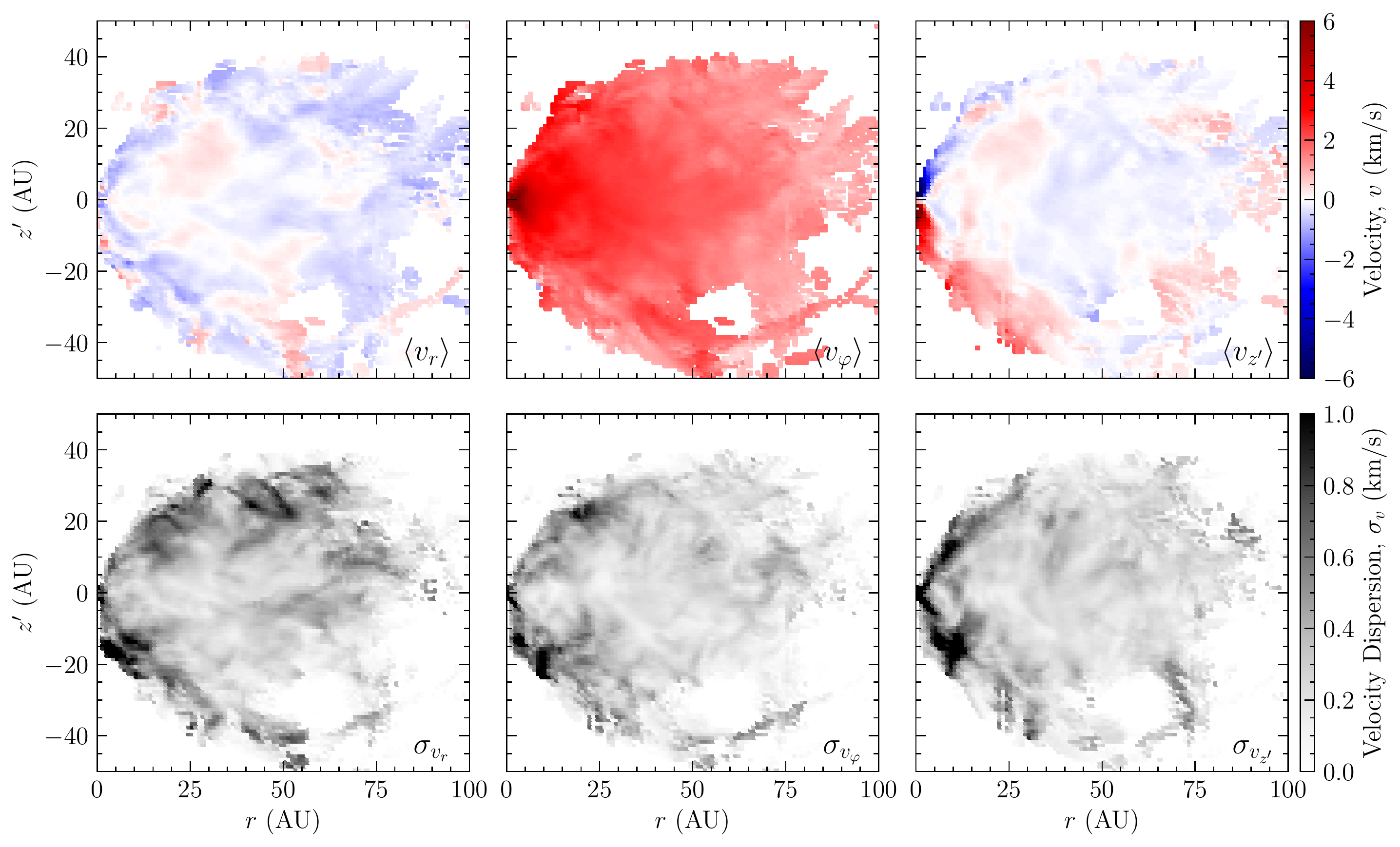}
    \caption{Mean velocity component map of $\langle v_r \rangle$ (top left), $\langle v_\varphi \rangle$ (top middle), and $\langle v_{z'} \rangle$ (top right); and their respective dispersions $\sigma_{v_r}$ (bottom left), $\sigma_{v_\varphi}$ (bottom middle), and $\sigma_{v_{z'}}$ (bottom right) in the $r$-$z'$ plane of the disc at $\tau=\kyr{10}$. The peaks in the magnitudes of the velocity components are concentrated near the centre, suggesting steady accretion through most of the disc, with increased kinematic activity close to the sink particle. The $\langle v_\varphi \rangle$ panel shows the systematic rotation, and $\langle v_{z'} \rangle$ indicates inflow of gas along the $z'$-axis near the sink, although we also see evidence of out-flowing material in regions at $r>\AU{10}$. The dispersion maps (bottom panels) indicate significant kinematic variability near the central region, i.e., driving of turbulence by the in-flowing material, creating shearing motions, and turbulence activity at a somewhat lower level throughout the disc.}
    \label{fig:mapVs}
\end{figure*}

For a more quantitative view of the kinematics, with the disc geometry in a cylindrical coordinate system, Fig.~\ref{fig:mapVs} shows the mean velocities (top row) and velocity dispersions (bottom row) in the $r$-$z'$ plane at $\tau=\kyr{10}$ by averaging $v_r$, $v_\varphi$, and $v_{z'}$ over all $\varphi$ at a given $r$ and $z'$. 

First focussing on the mean velocities in the top panels of Fig.~\ref{fig:mapVs}, we find that the highest magnitudes of velocity are in $v_\varphi$, quantifying the disc rotation, and in $v_z'$ near the sink particle, indicating in-flow near the sink particle.

In the $v_r$ map, there is no clear indication of large-scale structures, with small positive and negative components relatively evenly distributed across the $r$-$z'$ plane. As radial velocity tracks material being brought in from the outer disc, we indeed see a slight dominance of negative radial velocities in and near the disc mid-plane, indicating accretion flows through the disc. However, there is significant disturbance and variation in $v_r$, supporting our earlier observation of erratic and episodic accretion due to the constant turbulence present in the simulation.

The $v_\varphi$ map shows high velocities around the sink particle ($v_\varphi \sim \kms{10}$), indicating rotational motion speeding up towards smaller radii due to the conservation of angular momentum, i.e., $v_\varphi$ shows Keplerian to slightly sub-Keplerian motion inside the disc (see following Sec.~\ref{sec:keplerian} for a more detailed analysis).

The $v_z'$ map shows bifurcation of $v_z'$ about the mid-plane ($v_z'<0$ for $z'>0$ and $v_z'>0$ for $z'<0$) near the sink particle ($r\lesssim\AU{10}$), indicating inflow. However, for $r>\AU{10}$, we see regions with out-flowing portions, e.g., $v_z'>0$ for $z'>0$ in the range $10\lesssim r/\mathrm{AU}\lesssim30$, and $v_z'<0$ for $z'<0$ in the range $30\lesssim r/\mathrm{AU}\lesssim80$. It should be noted, however, that these flows are subject to highly intermittent fluctuations, with episodes of accretion and outflow switching locally on timescales of $\sim100\,\mathrm{yr}$, similar to the overall episodic accretion analysed in Sec.~\ref{sec:12evo}. Thus, while there is some evidence of out-flowing material, we do not find a coherent jet, which may be attributable to the turbulent magnetic field configuration associated with the highly dynamic turbulent flows, suppressing coherent jet launching \citep{GerrardFederrathKuruwita2019}. Nevertheless, at later stages of the evolution we may expect some settling of the disc and turbulent dynamics, potentially allowing for a jet to form, which would ultimately break out of the core \citep{TafallaMyers1997,ArceGoodman2001,StojimirovicEtAl2006,ArceEtAl2010,NarayananSnellBemis2012,FederrathEtAl2014,Federrath2015,MathewFederrath2021}.

In addition to the systematic motions that these averages over $\varphi$ represent, we also quantify the variations in $\varphi$, by calculating the velocity dispersion, $\sigma_v$, as
\begin{equation} \label{eqn:sigmaV}
    \sigma_{v_j} = (\langle v_j^2 \rangle_\mathrm{mw} - \langle v_j \rangle_\mathrm{mw}^2)^{1/2},
\end{equation}
where $j = \{r, \varphi, z'\}$ are the three components, and $\langle v_j \rangle_\mathrm{mw}$ is the mass-weighted mean velocity of component $j$ of all cells at a given $r$ and $z'$.

The velocity dispersions are shown in the bottom panels of Fig.~\ref{fig:mapVs}. In these maps, we see similar levels of dispersion with $\sigma_v\sim 0.3-\kms{0.6}$ across all three components throughout the disc, indicating nearly isotropic turbulence, despite the systematic large-scale motions of rotation of the disc. An exception to the isotropy is the enhanced dispersion in $\sigma_{v_{z'}}$ of up to $\sim\kms{1}$ near the sink particle, where we found strong inflows (c.f., top right panel at $r\lesssim\AU{10}$). Thus, this enhanced dispersion is likely instigated by the sheering motion of the gas inflow, with Kelvin–Helmholtz instabilities (KHI) forming. In the $\sigma_{v_\varphi}$ map, the dispersion around the inflow can be attributed to the non-linear interaction between the velocity components, as KHI creates perturbations that become unstable and form vortices, developing dispersion in other directions. In the envelope regions of the disc, we see additional, non-trivial dispersion, likely indicating the infall of gas from the progenitor core, creating turbulence while being accreted onto the disc.

\begin{figure*}
    \centering
    \includegraphics[width=\linewidth]{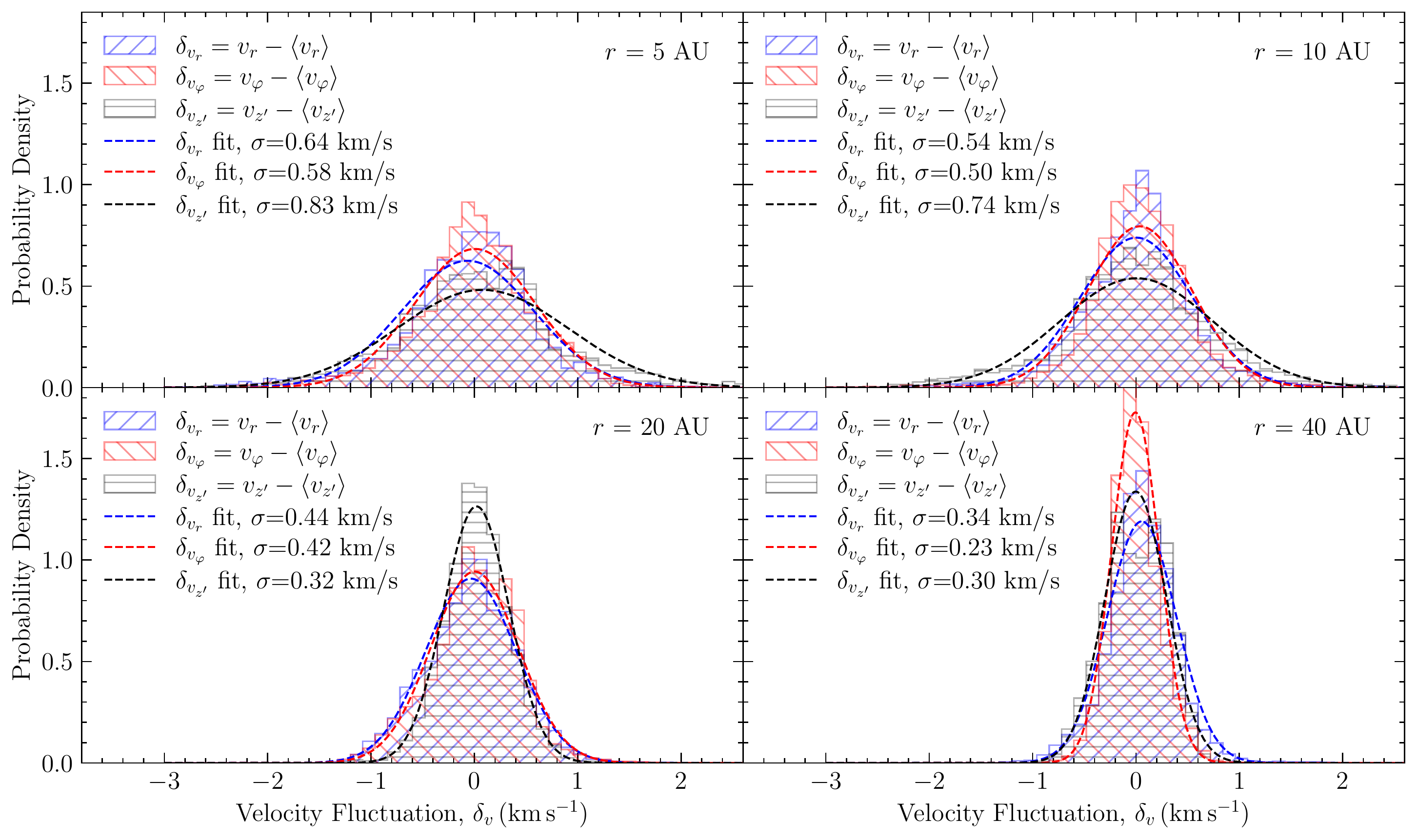}
    \caption{Probability distribution functions of the fluctuation components of the velocities (histograms), and fitted Gaussian distributions (dashed lines) at $r=5$, $10$, $20$, and $\AU{40}$ for the disc at $\tau=\kyr{10}$. The fluctuations decrease in all velocity components as $r$ increases. The inflow motion (c.f., top right panel of Fig.~\ref{fig:mapVs}) is significant at $r\sim5-\AU{10}$, increasing $\delta_{v_{z'}}$ compared to the other two components.}
    \label{fig:flucVMap}
\end{figure*}

\begin{figure}
    \centering
    \includegraphics[width=\linewidth]{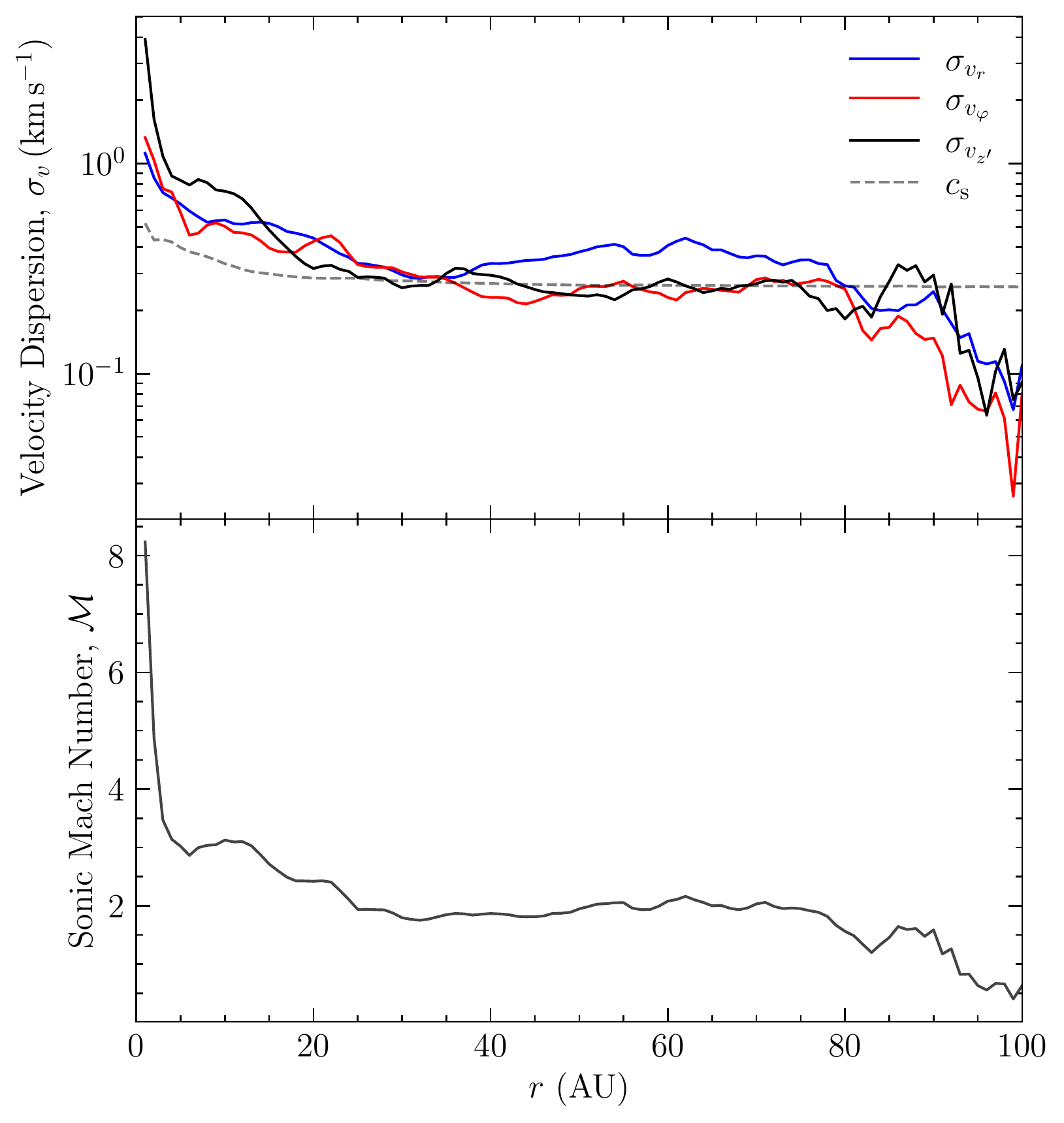}
    \caption{Velocity dispersion components and sound speed (top) and corresponding sonic Mach number (bottom) as a function of disc radius. The velocity dispersion components are largely of the order of the local sound speed for most radii, but differences start to emerge closer to the sink particle, where the inflow (c.f., top right panel of Fig.~\ref{fig:mapVs}) causes supersonic motions.}
    \label{fig:vDispersionProfile}
\end{figure}

\subsubsection{Turbulent velocity fluctuations} \label{sec:turbV}

Using Fig.~\ref{fig:mapVs} as a `lookup table', we subtract the mean velocity values (top panels) from each computational cell, determined by its $r$ and $z'$ positions. This leaves the fluctuating components of the respective velocity component.

Fig.~\ref{fig:flucVMap} shows the probability distribution functions (PDFs) and fitted Gaussian distributions of the velocity fluctuations at $r=5$, $10$, $20$, and $\AU{40}$ for the disc at $\tau=\kyr{10}$. Overall, the PDFs follow a Gaussian distribution, which is characteristic of turbulent flow \citep{KritsukEtAl2007,Federrath2013}. We observe a decrease in fluctuations as $r$ increases for all velocity components, seen as the narrowing of the width of the PDFs. This is caused by the higher turbulent environment in the centre close to the sink particle, where the fast inflow induces additional dispersion, i.e., at $r\sim5-\AU{10}$, $\delta_{v_{z'}}$ has a somewhat larger dispersion (by about $40-50\%$) than the other two components due to influence from the inflow along the $z'$-axis. The dispersions of $\delta_{v_r}$ and $\delta_{v_\varphi}$ also increase due to the non-linear interaction between components in the KHI vortices close to the sink particle.

We calculate the velocity dispersion along the entire disc using Equation~(\ref{eqn:sigmaV}) to produce a radial profile of $\sigma_v$. The result is shown in the top panel of Fig.~\ref{fig:vDispersionProfile}. The dispersion components are largely of the order of the local sound speed at $\AU{25} \lesssim r \lesssim \AU{75}$. Closer to the sink particle, at $r \lesssim \AU{10}$, the dispersions increase due to influence of the inflow; at $r \gtrsim \AU{75}$ the dispersions fall off as the disc dissipates. We then use the 3D velocity dispersion,
\begin{equation} \label{eqn:sigmaV3D}
    \sigma_v = \left( \sigma_{v_r}^2 + \sigma_{v_\varphi}^2 + \sigma_{v_{z'}}^2 \right)^{1/2},
\end{equation}
to produce a radial profile of the sonic Mach number, $\mathcal{M} = \sigma_v/\cs$. The result is shown in the bottom panel of Fig.~\ref{fig:vDispersionProfile}. Similar to the trend of $\sigma_v$, the sonic Mach number is roughly constant ($\mathcal{M} \sim 2$) at $\AU{25} \lesssim r \lesssim \AU{75}$, and increases closer to the sink particle due to the supersonic motions driven by the inflow.

\subsection{Disc magnetic field statistics} \label{sec:magnetics}

\subsubsection{Structure and morphology}
\label{sec:BMorphology}

In the previous Fig.~\ref{fig:slice}, we also show blue magnetic field streamlines. The face-on view (left panel) shows magnetic fields being wound up by the rotational motion of the disc, as evident in the concentric circular streamlines about the centre, extending out to $r\sim\AU{70}$. This is somewhat larger than the estimated radius of the disc from Fig.~\ref{fig:12_time_evo}. Compared to the uniform rotational velocity field, we see somewhat more variable structures since more streamlines enter and exit the field of view without looping around the centre.

The edge-on view (right panel) reveals the bipolar magnetic field `rails' above and below the sink particle along the $z'$-axis, which are responsible for the magneto-centrifugal launching of material seen in the $\langle v_{z'} \rangle$ panel of Fig.~\ref{fig:mapVs}. Inside the disc itself, we find a more random distribution of field lines with a number of small-scale ($r\sim\AU{10}$) closed loops, again due to the high level of turbulence in the disc.

\subsubsection{2D maps of magnetic field strength} \label{sec:Bmaps}

\begin{figure*}
    \centering
    \includegraphics[width=\linewidth]{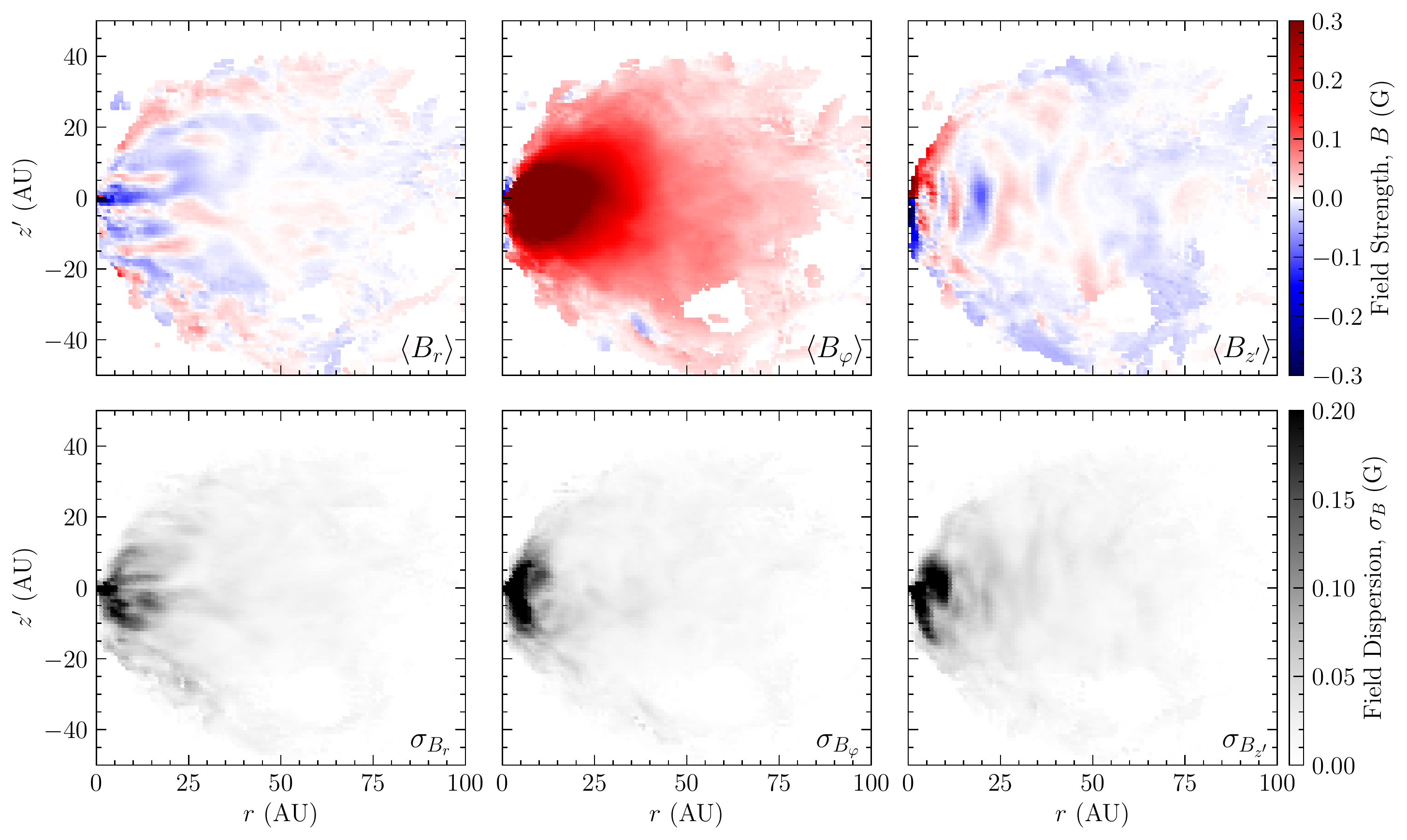}
    \caption{Mean magnetic field strength component map of $\langle B_r \rangle$ (top left), $\langle B_\varphi \rangle$ (top middle), and $\langle B_{z'} \rangle$ (top right); and their respective dispersions $\sigma_{B_r}$ (bottom left), $\sigma_{B_\varphi}$ (bottom middle), and $\sigma_{B_{z'}}$ (bottom right) in the $r$-$z'$ plane of the disc at $\tau=\kyr{10}$. The peaks in the magnitudes of the field strength components are concentrated near the centre, dominated by $B_\varphi$, suggesting winding up of magnetic field from the rotational motion of the disc. Field strength dispersions are localised around the sink particle, due to the perturbations induced by the inflow.}
    \label{fig:mapBs}
\end{figure*}

Again with the disc geometry in a cylindrical coordinate system, and similar to Fig.~\ref{fig:mapVs}, Fig.~\ref{fig:mapBs} shows the mean magnetic field strength (top row) and field strength dispersions (bottom row) in the $r$-$z'$ plane at $\tau=\kyr{10}$ by averaging $B_r$, $B_\varphi$, and $B_{z'}$ over all $\varphi$ at a given $r$ and $z'$. Overall, the highest magnitudes of magnetic field are found near the sink particle in each case, with $B_\varphi$ dominating due to the winding-up of the magnetic field in the disc. Compared to the disc at sink particle formation, the mean field strength has increased significantly, by $\sim2$~orders of magnitude for $B_r$ and $B_{z'}$, and by $\sim4$~orders of magnitude for $B_\varphi$.

The magnetic field dispersion, $\sigma_B$, is calculated from the field strengths at a given $r$ and $z'$,
\begin{equation} \label{eqn:sigmaB}
    \sigma_{B_j} = (\langle B_j^2 \rangle - \langle B_j \rangle^2)^{1/2},
\end{equation}
where $\langle B_j \rangle$ and $\langle B_j^2 \rangle$ are respectively the volume-weighted mean and mean-squared magnetic field components ($j = \{r, \varphi, z'\}$) of all cells at a given $r$ and $z'$.

In the $B_r$ map, we observe the region of high $B_\varphi$ around the centre, spanning $\sim\AU{20}$ in both $r$ and $z'$-directions, which is a result of the magnetic field being wound up by the rotational motion of the disc. Around the centre, we see negative $B_r$ along the disc mid-plane, indicating the magnetic field lining up with the radially inward accretion flow and amplifying closer to the sink particle. The distribution of $B_{z'}$ again shows the magnetic `rails' above and below the sink particle along which the disc material is launched.

In the $\sigma_B$ maps, we see the field strength dispersions are localised around the sink particle, roughly constrained to the immediate regions near the inflow. Compared to the velocity dispersions in the bottom row of Fig.~\ref{fig:mapVs}, the magnetic field dispersions lack any significant structures outside the inflow region since the magnetic field, likely because of the relatively strong $B_\varphi$ component, which is hard to perturb, except through the fast motion of the inflow. We also notice the presence of magnetic `bubbles' through `interchange instability' \citep[see][]{Vaytet_2018} at larger radii ($r\gtrsim\AU{10}$). We note that these effects are expected to be less efficient or absent in non-ideal MHD (see also Sec.~\ref{sec:limitations} below).

\begin{figure*}
    \centering
    \includegraphics[width=.95\linewidth]{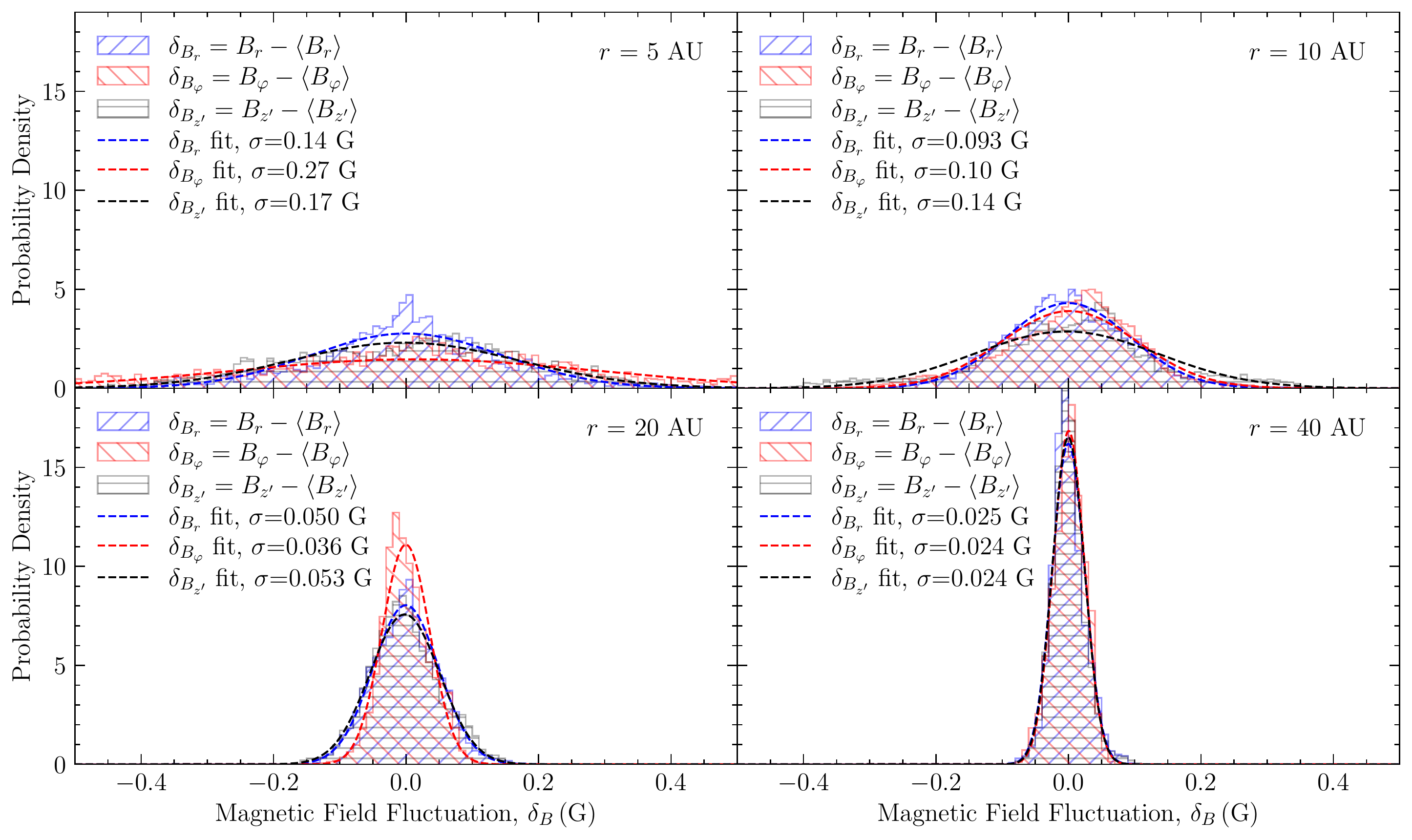}
    \caption{Probability distribution functions of the fluctuating components of the magnetic field (histograms) and fitted Gaussian distributions (dashed lines) at $r=5$, $10$, $20$, and $\AU{40}$ for the disc at $\tau=\kyr{10}$. The $B$ field fluctuations decrease in all components as $r$ increases, similar to the velocity dispersion (c.f., Fig.~\ref{fig:flucVMap}). Significant effects from the inflow at $r=5-\AU{10}$ widen the PDFs and drive magnetic field perturbations.}
    \label{fig:flucBMap}
\end{figure*}

\begin{figure}
    \centering
    \includegraphics[width=\linewidth]{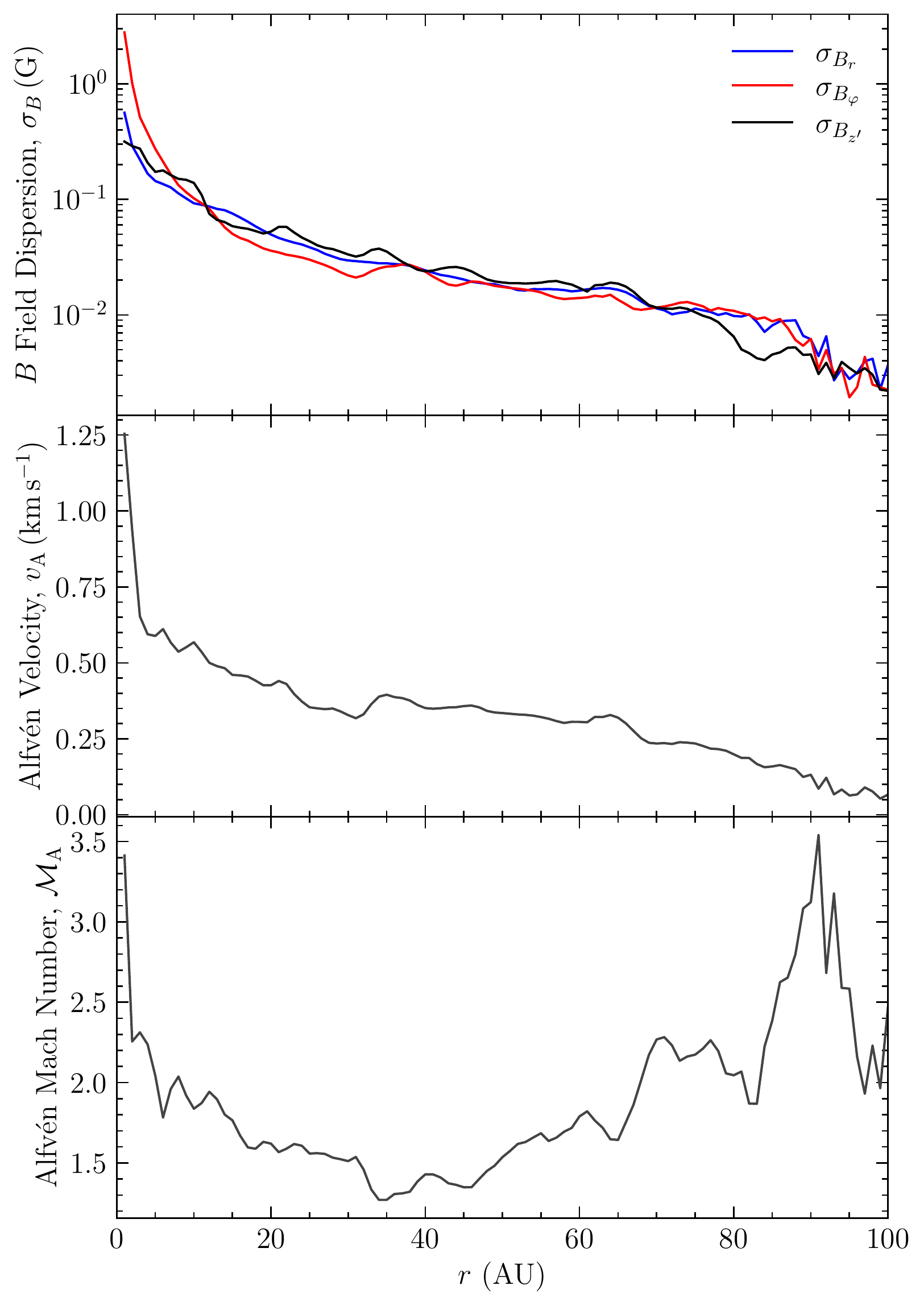}
    \caption{Magnetic field dispersions (top), Alfv\'en velocity (middle), and corresponding Alfv\'en Mach number (bottom) as a function of disc radius. A strong influence from the inflow can be seen in all quantities close to the sink particle. The turbulent Alfv\'en Mach number is $\sim2$, indicating a significant influence of magnetic perturbations on the internal dynamics of the disc and vice versa.}
    \label{fig:BDispersionProfile}
\end{figure}

\subsubsection{Turbulent magnetic field fluctuations}

Following the same procedure as described in Section~\ref{sec:turbV}, we produce Fig.~\ref{fig:flucBMap} from the fluctuating components of the magnetic field, showing the PDFs and fitted Gaussian distributions of the field strength fluctuations at $r=5$, $10$, $20$, and $\AU{40}$ for the disc at $\tau=\kyr{10}$. The PDFs are well described by Gaussian distributions. We observe a decrease in the fluctuations as $r$ increases for all components, seen as the narrowing of the width of the PDFs -- caused by the turbulent environment in the centre, increasing fluctuations due to the presence of the inflow.

We calculate the magnetic field dispersion along the entire disc using Equation~\ref{eqn:sigmaB} to produce a radial profile of $\sigma_B$. The result is shown in the top panel of Fig.~\ref{fig:BDispersionProfile}. We see a steady decrease in the dispersions as $r$ increases throughout the entire disc, but closer to the sink particle, at $r \lesssim \AU{10}$, the dispersions increase due to influence from the inflow. Beyond $r \gtrsim \AU{75}$, disc dissipation makes the dispersion profiles erratic. We then use the 3D velocity dispersion (Eq.~\ref{eqn:sigmaV3D}) and 3D magnetic field dispersion,
\begin{equation} \label{eqn:sigmaB3D}
    \sigma_B = \left( \sigma_{B_r}^2 + \sigma_{B_\varphi}^2 + \sigma_{B_{z'}}^2 \right)^{1/2},
\end{equation}
to produce a radial profile of the Alfv\'en speed, $\va=\sigma_B/\left(4\pi\rho\right)^{1/2}$ and the corresponding Alfv\'en Mach number, $\macha=\sigma_v/\va$. The results are shown in the middle and bottom panels of Fig.~\ref{fig:BDispersionProfile}. The Alfv\'en velocity follows a similar trend of decay as the dispersions with increasing $r$, whereas the Alfv\'en Mach number initially decreases to a minimum of $\macha\sim1-1.5$ at $r \sim \AU{40}$, and increases towards smaller and larger radii from there, reaching values as high as $\macha\sim3.5$. Thus, since $\macha\sim1-3$ overall, the magnetic field has a substantial impact on the turbulent density and velocity fluctuations in the accretion disc \citep{MolinaEtAl2012,BeattieFederrathSeta2020}. Combining our measurements of the sonic and Alfv\'en Mach number, we can estimate the turbulent component of plasma beta, i.e., the ratio of thermal to magnetic pressure in the disc as \citep{FederrathKlessen2012}
\begin{equation}
\beta = 2 \cs^2 / \va^2 = 2 \macha^2 / \mach^2,
\end{equation}
which gives $\beta \sim 1$ for the disc, i.e., the thermal and turbulent magnetic pressure are comparable. We note that upon inclusion of non-ideal MHD effects we would expect $\beta$ to be somewhat higher, by factors of a few \citep{Masson_2016} -- see also the discussion on non-ideal MHD in Sec.~\ref{sec:limitations} below.

\subsubsection{Alfv\'en surface analysis}

\begin{figure*}
    \centering
    \centering
    \includegraphics[width=.95\linewidth]{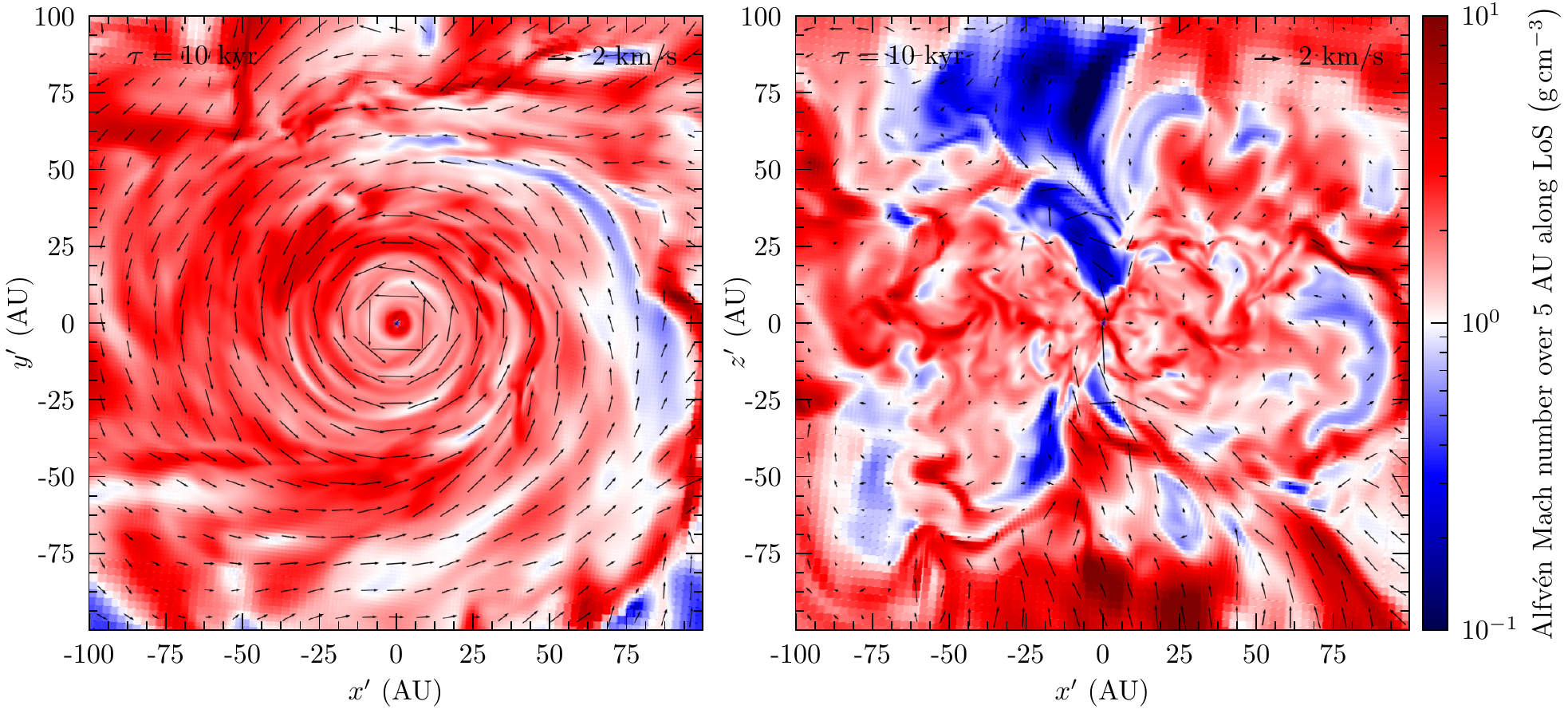}
    \caption{Same as Fig.~\ref{fig:slice}, but for the Alfv\'en Mach number. Black arrows represent velocity vectors of the gas motion. White coloured regions represent Alfv\'en surfaces ($\macha=1$). The disc plane is primarily characterised by $\macha\gtrsim1$, with some regions above and below the disc plane showing $\macha\lesssim1$. An animation (see online version) reveals intermittent outflows in the form of rising magnetic structures (`bubbles').}
    \label{fig:alfven}
\end{figure*}

To evaluate the importance of the magnetic field relative to gas flows in local regions of the disc (in the midplane as well as above and below), we plot the Alfv\'en Mach number, $\macha$ in Fig.~\ref{fig:alfven} (similar to Fig.~\ref{fig:slice}). In the disc mid-plane, we find super-Alfv\'enic conditions ($\macha\gtrsim1$), because of the fast rotational speeds and relatively low $\va$ (noting that both $B$ and $\rho$ are higher in the disc than in the outskirts). However, above and below the mid-plane, we see some sub-Alfv\'enic regions, where the magnetic field is dominating the fluid motions. An animation is provided in the online version, showing that some of these regions intermittently rise as `bubbles' from the disc (as briefly mentioned at the end of Sec.~\ref{sec:Bmaps}), i.e., represent out-flowing material. However, this is a highly episodic process, where flows are complex and change from out-flowing to in-flowing on timescales of $\sim100\,\mathrm{yr}$ (c.f., similar to the episodic accretion timescales discussed in Sec.~\ref{sec:12evo}).

\section{Limitations and future work} \label{sec:limitations}

\subsection{Radiation transport}

Radiation transport is important in shaping the thermal structure of a protostellar disc. Accurate radiation transport modelling, however, remains challenging, as it ideally involves time-dependent, multi-wavelength solutions at high angular resolution of the radiation transport equations \citep[e.g.,][]{Harries_2017}. Most MHD simulations adopt approximations, e.g., using a polytropic equation of state as we have done here (see Sec.~\ref{sec:EoS}), to manage computational costs. This approach captures the overall thermal evolution, but neglects radiation feedback, possibly underestimating local temperature gradients and their associated effects on accretion dynamics \citep{OffnerEtAl2009}. Improved radiation-hydrodynamics models \citep[e.g.,][]{MenonEtAl2022} should be considered in future studies using the re-simulation technique.

\subsection{Radiation feedback}

As the protostar accretes material, gravitational energy converts to radiation, heating the surrounding gas \citep{MathewFederrath2020}. This heating creates a pressure gradient that affects accretion rates, potentially contributing to the episodic accretion patterns seen in young protostellar systems. Radiation feedback may therefore significantly influence the structure and evolution of the disc. The feedback can also influence the presence and dynamics of jets, outflows, and magnetic `bubbles'. Incorporating radiative feedback methods would enhance our ability to simulate more realistic varied disc structures and accretion behaviours, which needs to be explored in future studies.

\subsection{Numerical resolution}

In this study, our simulation achieves a maximum effective resolution of $\AU{0.63}$, sufficient to capture large-scale disc dynamics, but still limited in resolving finer-scale structures within the protostellar disc. We use AMR to distribute computational resources based on Jeans refinement (see Sec.~\ref{sec:AMR}). However, this means lower density regions are not simulated to the same resolution, similar to smoothed particle hydrodynamics (SPH) methods \citep{Price2012}. However, in contrast to our grid-based approach, SPH methods such as those explored by \citet{BateTriccoPrice2014} offer adaptable resolution that can often resolve smaller-scale interactions more flexibly.

Going beyond the base criterion for resolving gravitational collapse without artificial fragmentation, we implemented a Jeans resolution criterion (see Sec.~\ref{sec:jeans}) enforcing a minimum resolution of 30~cells per Jeans length, which is critical for accurately capturing the disc structure as well as turbulent and magnetic dynamics at small scales \citep[][fig.~11]{FederrathSurSchleicherBanerjeeKlessen2011,FederrathEtAl2014}. Many state-of-the-art simulations use a lower Jeans resolution, which, while allowing for a higher maximum resolution, may not be sufficient to capture the details of the turbulent dynamics and magnetic field structures in the disc. Future simulations would need to resolve both the Jeans length with $\gtrsim30$~cells throughout, and push the boundaries in terms of maximum resolution.

\subsection{Non-ideal MHD}

Our simulations only employ ideal MHD. Non-ideal MHD effects (ambipolar diffusion, Ohmic dissipation, Hall effect) may be highly relevant for the disc structure and dynamics, in particular in the higher-density regions of the disc, where low ionization levels render the ideal MHD approximation exceedingly inaccurate \citep{Masson_2016,NolanEtAl2017,Wurster_2018,Vaytet_2018,Zhao_2020,TritsisEtAl2022,Tsukamoto_2023,Kuffmeier_2024,Mayer_2025}. For instance, these effects (often dominated by ambipolar diffusion) can hinder but not stop the conversion of turbulent energy to magnetic energy, i.e., dynamo amplification or at least maintenance of magnetic fields, which may not only be relevant in accretion discs around the first stars \citep{SchoberEtAl2012,NakauchiOmukaiSusa2019,McKeeStacyLi2020,ShardaEtAl2021,Sadanari_2023}. However, close to the protostar, where ionizing feedback becomes relevant, the role of non-ideal MHD is not yet fully understood and warrants further investigation. Finally, we note that the disc scale height discussed in Sec.~\ref{sec:scale_height} may be overestimated due to the assumption of ideal MHD \citep{Masson_2016}.

\subsection{Duration of disc evolution}

The simulation in this study was evolved to $\kyr{10}$ after sink particle formation, capturing the initial formation and growth of the protostellar disc. Extending the simulation duration would allow us to study the longer-term structural and dynamical evolution within the disc including potential development of more stable spiral arms, gaps, as well as the impact of continued episodic accretion events in the turbulent environment on the disc’s density, kinematics, and magnetic field evolution. However, the absence of radiation feedback (as discussed previously) limits our ability to simulate fully realistic systems. Moreover, longer time evolution is required to capture potential disc fragmentation \citep[e.g.,][]{Kuruwita_2024}.

\section{Conclusions} \label{sec:conclusions}

We presented a high-resolution MHD simulation of the formation and early evolution a young protostellar disc embedded in a turbulent, magnetised, and self-gravitating environment inherited from a $\left(\pc{2}\right)^3$ molecular cloud simulation. We simulated the star+disc system in a $\left(\pc{0.1}\right)^3$ domain to $\kyr{10}$ after sink/star formation, with a maximum effective resolution of $\AU{0.63}$. Our approach allowed us to capture protostellar disc formation and evolution under the influence of self-consistent initial conditions for the turbulence, magnetic fields, and mass distributions derived from the parent molecular cloud, which are critical ingredients in shaping the dynamical and structural properties of young discs.

Our key findings can be summarised as follows:

\begin{enumerate}

    \item The initially dense core on the verge of collapse evolves into a rotationally supported protostellar disc over the first $\kyr{10}$ after sink particle formation. The disc grows to radii on the order of several tens of AU, with episodic accretion events driving fluctuations in both disc mass and radius. Despite these fluctuations, a general trend of growth persists, supported by ongoing accretion from the turbulent environment. At $\tau=\kyr{10}$, the disc reaches a mass of $\msol{0.12}$ around a $\msol{0.15}$ protostar.
    
    \item The disc's density profile steepens over time, transitioning from nearly flat at the onset of disc formation to a roughly power-law distribution with an exponent $\sim 1$ at $\tau=\kyr{10}$. We find the disc to be geometrically thick, with a flaring structure and scale height $H \sim 5-\AU{10}$ at radial distances of tens of AU. This substantial vertical extent partly reflects the significant turbulence and continuous accretion from the surrounding cloud at this early stage of the evolution.

    \item The surface density profile is somewhat shallower than the minimum mass solar nebula (MMSN), but steepens over time, such that our simulations is consistent with the expected MMSN for solar system formation.

    \item The accretion onto both the protostar and the disc is highly episodic, with brief bursts of enhanced mass flux of $\dot{m}\sim10^{-5}\,\mathrm{M_\odot\,yr^{-1}}$ followed by periods of reduced activity. Such episodic behaviours are consistent with observational evidence of variable protostellar accretion rates, as well as simulation results in the literature.

    \item The disc kinematics is strongly influenced by the inherited turbulence from the molecular cloud scale. The disc is mildly sub-Keplerian ($v_\varphi/\vk\sim0.8-0.9$), yet remained strongly turbulent. The velocity dispersions are generally of the order of the local sound speed, resulting in a turbulent sonic Mach number of $\mathcal{M} \sim 2$. At the interface between the rotating disc and inflowing gas, velocity shear generates Kelvin–Helmholtz instabilities, forming vortex-like structures that enhance local turbulence to sonic Mach numbers exceeding 3.

    \item The disc inherits a relatively weakly magnetised environment from the parental cloud, which is amplified through gravitational collapse and disc rotation. Near the sink particle, the field lines are wound up, resulting in a strong azimuthal field component and significant magnetic field dispersion. The turbulent field structures allow for intermittent magnetic `bubbles' to form and be transported away from the disc. However, coherent jet formation is suppressed due to the turbulent field configuration \citep[c.f.][]{GerrardFederrathKuruwita2019}.

    \item The turbulent Alfv\'en Mach number of the disc is $\sim2$, and the plasma beta is $\sim1$, indicating a significant influence of the magnetic field on the dynamics and density structure of the disc, likely impacting the ability of the disc to fragment despite the presence of strong density perturbations including spiral-arm features during its early evolution.

\end{enumerate}

\section*{Acknowledgements}
The authors thank the anonymous referee for providing valuable insights and helpful suggestions, which improved the work. C.~F.~acknowledges funding provided by the Australian Research Council (Discovery Project grants~DP230102280 and~DP250101526), and the Australia-Germany Joint Research Cooperation Scheme (UA-DAAD). We further acknowledge high-performance computing resources provided by the Leibniz Rechenzentrum and the Gauss Centre for Supercomputing (grants~pr32lo, pr48pi, pn76ga and GCS Large-scale project~10391), the Australian National Computational Infrastructure (grant~ek9) and the Pawsey Supercomputing Centre (project~pawsey0810) in the framework of the National Computational Merit Allocation Scheme and the ANU Merit Allocation Scheme. The simulation software, \texttt{FLASH}, was in part developed by the Flash Centre for Computational Science at the University of Chicago and the Department of Physics and Astronomy at the University of Rochester.

\section*{Data Availability}

The simulation data underlying this paper, which is a total of $\sim10\,\mathrm{TB}$, or specific parts thereof, will be shared on reasonable request to the authors.



\newcommand{\rmp}{Reviews of Modern Physics}
\bibliographystyle{mnras}
\bibliography{ref,federrath}




\appendix

\section{Re-simulation box size study} \label{appx:box_size}

Our main re-simulation is run in a cubic computational domain with side length $L=\pc{0.1}$. To investigate the dependence on the choice of $L$, seven re-simulations are run with $L=0.2$, $0.1$, $0.05$, $0.025$, $0.0125$, $0.00625$, and $\pc{0.003125}$. These re-simulations have the same maximum effective resolution of $\dx=\AU{2.52}$, i.e., a sink particle radius of $\rsink=2.5\,\dx=\AU{6.3}$.

\begin{figure}
    \centering
    \includegraphics[width=\linewidth]{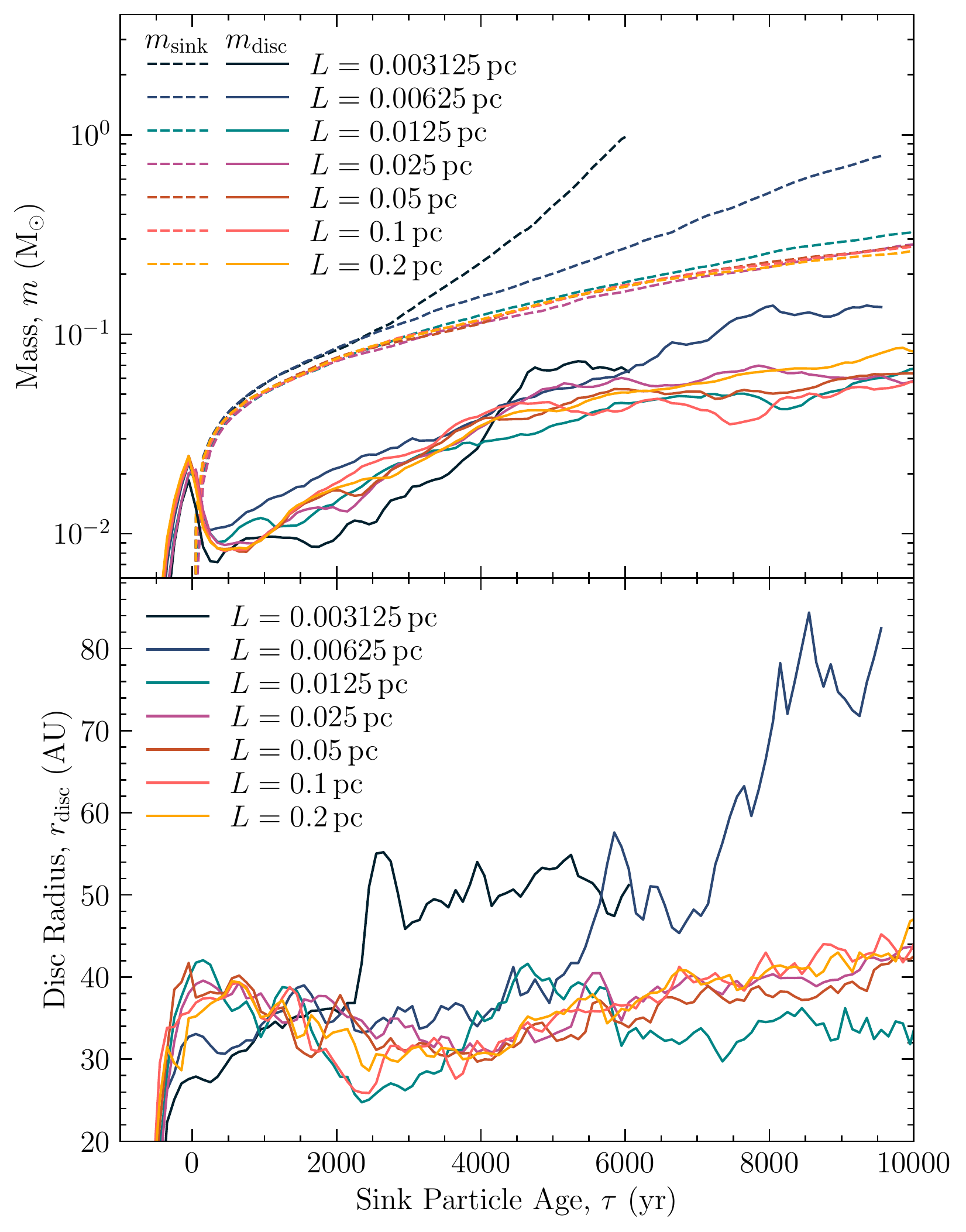}
    \caption{Time evolution of sink particle mass and disc mass (top), and disc radius (bottom) in simulations with different re-simulation box sizes, $L$, but with identical maximum effective resolution of $\dx=\AU{2.52}$. We find converged results for box sizes $L\ge\pc{0.025}$.}
    \label{fig:box_mass_size_evo}
\end{figure}

Fig.~\ref{fig:box_mass_size_evo} shows the time evolution of the sink particle and disc mass (top panel), and the disc radius (bottom panel). We see that for very small $L<\pc{0.025}$, the results vary substantially as the region from which the disc+star system accretes over the course of the simulation is $\sim\pc{0.025}$, and therefore the results are significantly affected by the boundary conditions of the re-simulation box when $L<\pc{0.025}$. In contrast, all results are converged for $L\ge\pc{0.025}$.

\section{AMR block structure and refinement fraction} \label{appx:amr}

\begin{figure*}
    \centering
    \includegraphics[width=\linewidth]{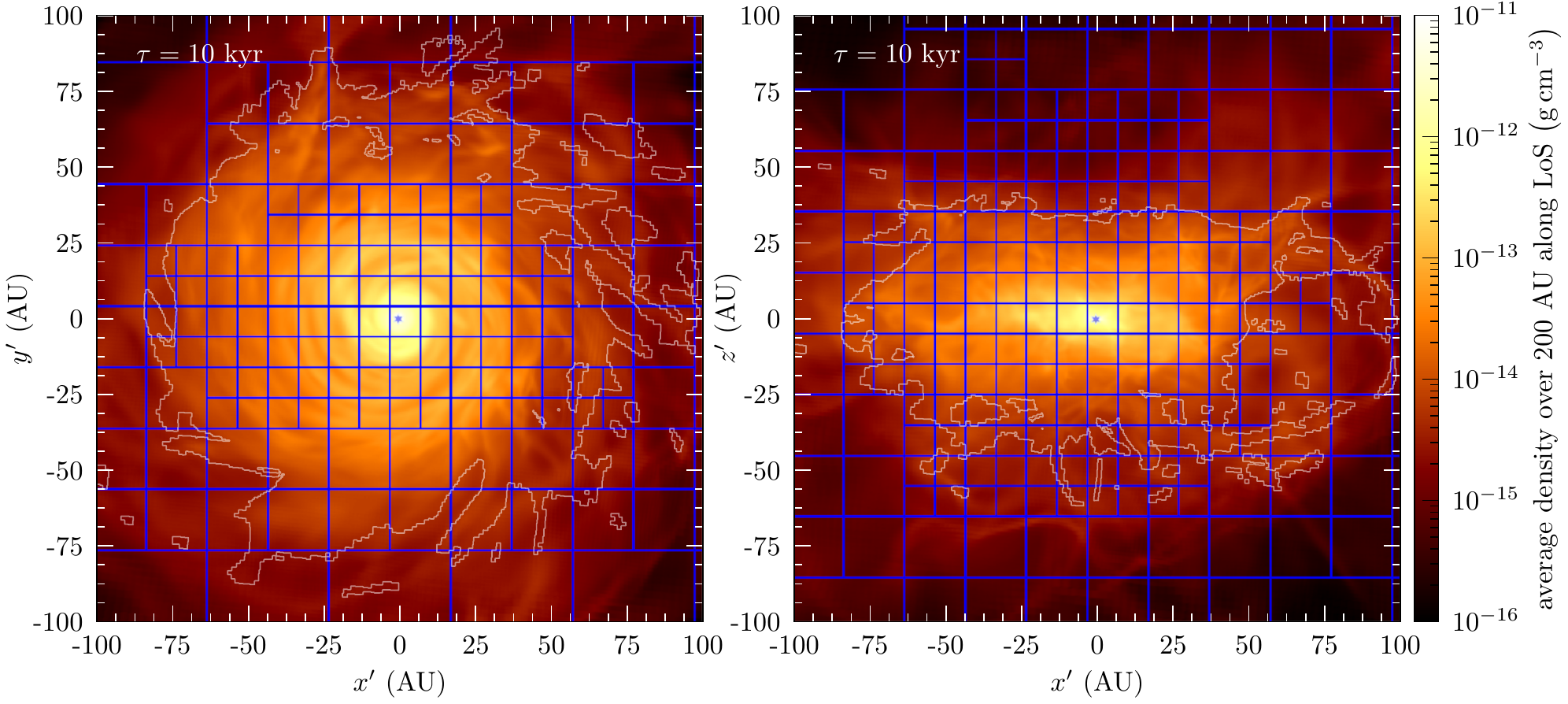}
    \caption{Same as the bottom panels of Fig.~\ref{fig:evo_rotation}, but with the AMR block structure superimposed as blue rectangles. Each block (rectangle) contains $16^3$~cells, i.e., has a side length of $16\,\dx$. The highest-level cells ($\dx=0.63\,\mathrm{AU}$) cover $\sim90\%$ of the disc volume.}
    \label{fig:amr}
\end{figure*}

Figure~\ref{fig:amr} shows the AMR block structure (blue rectangles) together with the disc material contours, as in the bottom panels of Fig.~\ref{fig:evo_rotation}. The refinement fractions of the disc are: 90\% resolved at the highest level ($\dx=\AU{0.63}$) and 10\% at $2\,\dx$ by volume (or 98\% resolved at $\dx$ and 2\% at $2\,\dx$ by mass). Thus, a major fraction of the disc is resolved at the maximum level of AMR.

\section{Numerical resolution study} \label{appx:reso}

Our main re-simulation has a maximum effective resolution of $\dx=\AU{0.63}$. To investigate the resolution dependence of our results, two additional re-simulations are run with maximum effective resolutions of $\AU{1.26}$ and $\AU{2.52}$ to compare with our main re-simulation.

\begin{figure}
    \centering
    \includegraphics[width=\linewidth]{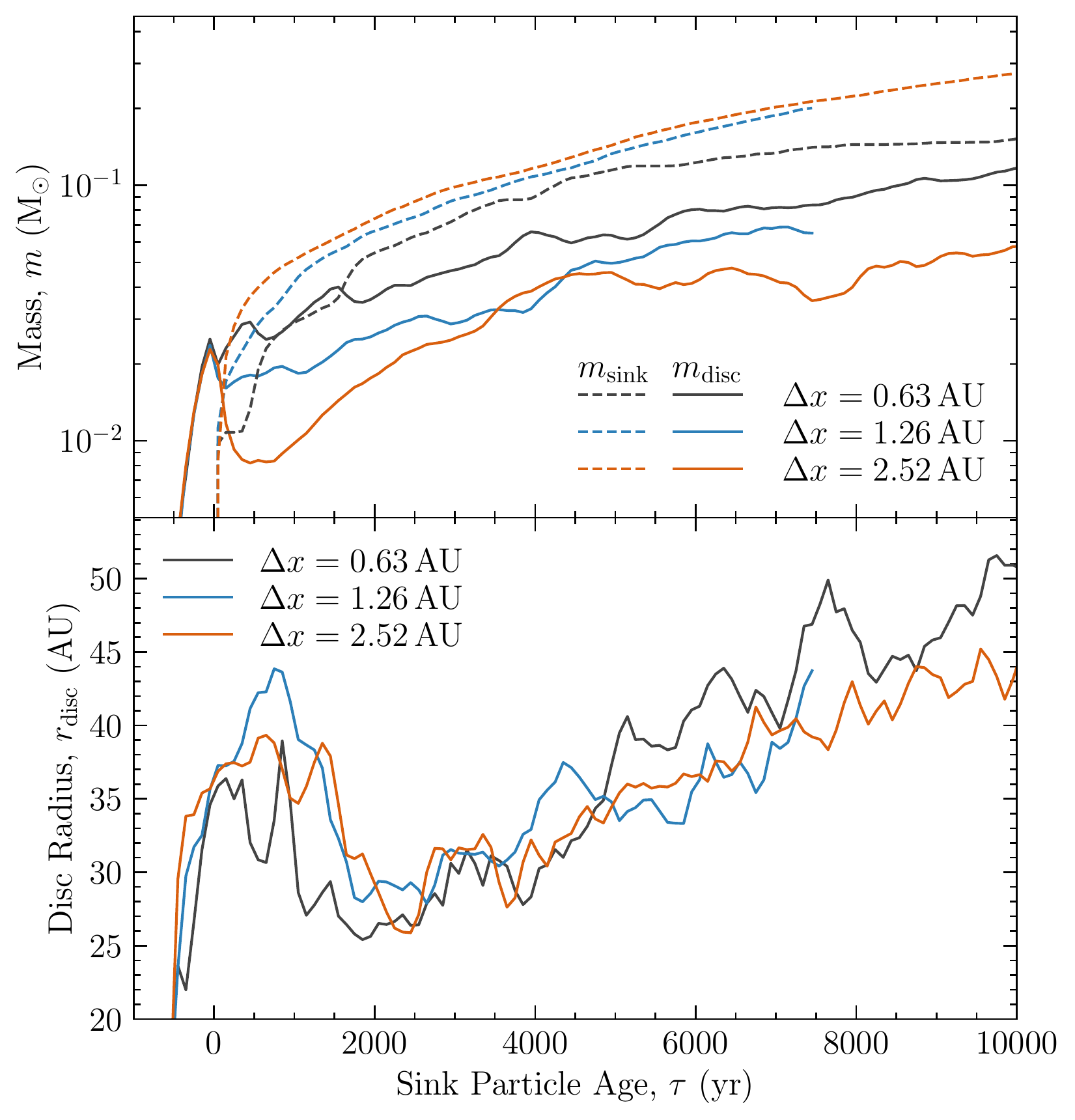}
    \caption{Time evolution of sink particle mass and disc mass (top), and disc radius (bottom), in simulations with different maximum effective resolution. All quantities show a converging trend with increasing resolution, i.e., decreasing $\dx$.}
    \label{fig:reso_mass_size_evo}
\end{figure}

Fig.~\ref{fig:reso_mass_size_evo} shows the same as Fig.~\ref{fig:box_mass_size_evo}, but for varying maximum effective resolution at fixed $L=\pc{0.1}$. Unsurprisingly, higher resolution produces smaller $\msink$ since the sink particle radius is determined by the maximum effective resolution, and smaller $\rsink$ encompasses less volume to accrete mass from. Naturally, this leaves more mass in the disc, resulting in higher $\mdisc$ for higher numerical resolutions. As the resolution increases, we find that the differences in $\mdisc$ and $\rsink$, progressively decrease from one resolution step to the next. Thus, the simulations are converging with increasing resolution, at least with respect to the disc properties that we are primarily interested in. However, we do see some indication of non-convergence with respect to the sink particle mass, $\msink$, and one would need to investigate a longer time evolution, as well as ideally adding lower- and high-resolution simulations, to provide a firm conclusion about the convergence behaviour.


\bsp	
\label{lastpage}
\end{document}